\documentclass[aps,pre,twocolumn,superscriptaddress,showpacs]{revtex4-1}

\usepackage{epsfig}
\usepackage{amsmath,amssymb,bm}
\usepackage{graphics}
\usepackage{epsfig}
\usepackage{graphicx}

\begin{document}

\title{
Core organization of directed complex networks}

\author{N. Azimi-Tafreshi} 

\affiliation{Departamento de F{\'\i}sica da Universidade de Aveiro $\&$ I3N, Campus Universit\'ario de Santiago, 3810-193 Aveiro, Portugal}

\author{S.~N. Dorogovtsev}
\affiliation{Departamento de F{\'\i}sica da Universidade de Aveiro $\&$ I3N, Campus Universit\'ario de Santiago, 3810-193 Aveiro, Portugal}
\affiliation{A.F. Ioffe Physico-Technical Institute, 194021 St. Petersburg, Russia}

\author{J.~F.~F. Mendes}
\affiliation{Departamento de F{\'\i}sica da Universidade de Aveiro $\&$ I3N, Campus Universit\'ario de Santiago, 3810-193 Aveiro, Portugal}



\begin{abstract} 
The recursive removal of leaves (dead end vertices) and their neighbors  
from an undirected network 
results, 
when this pruning algorithm stops, in a so-called core of the network. This specific subgraph should be distinguished from $k$-cores, which are principally different subgraphs in networks. 
If the vertex mean degree of a network is sufficiently large, the core is a giant cluster containing a finite fraction of vertices. 
We find that generalization of this pruning algorithm to directed networks provides a significantly more complex picture of cores. 
By implementing a 
rate equation approach to this pruning procedure for directed uncorrelated networks, 
we identify a set of cores progressively embedded into each other in a network and describe their birth points and structure.  
\\
\end{abstract}


\maketitle

\section{Introduction}

In terms of connectivity, a network, generally, is a set of connected components, which may include a giant connected component containing a finite fraction of vertices and edges (which is well-defined in infinite networks) \cite{struc1,struc2,Dorogovtsev:dm-books,Dorogovtsev:dgm08}. Studies of random graphs, starting from the first works \cite{Solomonoff:sr1951,giant1}, essentially focused on the connected components, their statistics and structure \cite{giant2}.
A natural generalization of connected components are $k$-cores, which are obtained by a special pruning procedure, namely the recursive removal of vertices with degrees smaller than $k$ \cite{seidman,Bollobas:b84}. The $k$-core structure of complex networks is being studied extensively \cite{kcore1,kcore2}. 
On the other hand, other key subgraphs of networks, 
so-called cores, are far less studied. For an undirected network, the core is a final result of the application of the following pruning algorithm. Remove recursively the leaves (by definition, vertices of degree $1$) with their nearest neighbors from the network. 
Speaking in more detail, at each step, remove a randomly chosen leaf and its neighboring vertex from the network. 
Continue the process until no leaves remain. 
  The resulting subgraph consists of finite components and, if the mean vertex degree $\langle q \rangle$ exceeds some threshold, of a giant component containing a finite fraction of vertices. It is this giant component that is usually called the core of a network \cite{Bauer}. The cores are related to a wide range of topical problems for networks such as controllability of networks
\cite{control}, localization on random graphs \cite{local}, and some
combinatorial optimization problems like maximum matching
\cite{matching,Hartmann:hrbook2001} and minimum vertex cover
\cite{vertexcover1,vertexcover2,Weight:whbook2005}. 

As a first application, this pruning 
algorithm was used by Karp and Sipser who intended to find a maximum matching in
graphs \cite{matching} (at least, some of the maximum matchings), where a matching is a set of edges of a graph which have no common vertices. They proposed an algorithm which recursively
selects an edge $(i,j)$ 
for which vertex $i$ has
degree $1$. This edge is included into the matching. 
Then all the edges incident to vertex $j$, including the edge $(i,j)$ are removed from the graph, 
and the process is recursively repeated. 
If the resulting graph has no leaves (e.g., when the graph is already reduced to the core) choose an edge at random, include it into the matching and remove all the edges incident to the end vertices of this edge. 
These steps are repeated until all the edges will be removed. The result of this procedure was suggested to be a maximum matching. 
Although it was shown that 
the Karp--Sipser 
algorithm, in general, actually does not generate the maximum  matchings \cite{Aronson:afp1998} 
it is often used as a preprocessing
step for other maximum matching algorithms \cite{Magun:m1998,Dietzfelbingerdpr:2012}.

For tree-like graphs, after the pruning, a remaining subgraph may consist only of isolated vertices and a giant core. 
Bauer and Golinelli studied the leaf removal algorithm on the Erd\H{o}s--R\'enyi graphs 
\cite{Bauer}, which are tree-like in the infinite network limit. Using generating
functions,  
they showed that there is a continuous phase transition at the mean degree $\langle
q\rangle = e$, where $e=2.718...$, so that below the transition the remaining subgraph
contains only isolated vertices 
and above the transition 
this subgraph consists of 
isolated
vertices and a giant 
core.

Previous studies of 
cores were mostly focused on
undirected networks, while numerous real-world networks, similarly to the World
Wide Web, are directed graphs (digraphs), i.e., their edges are directed. 
A leaf in directed networks is 
defined as a 
vertex of in- or out-degree equal to $1$, which gives in- or out-leaves respectively. 
The result of the recursive removal of these leaves (the algorithm is explained below in detail) is the core of a directed network \cite{control}.    
Importantly, there were found close relations between
the intensively studied controllability of directed networks and the emergence of a core and
the problem of maximum matching for directed graphs \cite{control}. These findings stimulated particular interest in the  
cores in directed networks. Recently, Liu {\em et al.} presented analytically derived 
results for cores  
in directed random graphs \cite{corenew}. 
They defined however the core in a different, peculiar way. They transformed a directed network into a bipartite graph and then found the core of this graph as if it was undirected unipartite. This convenient approach crucially simplifies the task and allows for a straightforward generating function technique but it actually leaves unsolved the original problem of the core structure of directed networks as it was defined in Ref.~\cite{control}. 

In this paper, 
we adapt the leaf removal algorithm to the case of digraphs. 
This 
extended leaf removal algorithm is 
introduced in such a way that it fits 
the definition of matching patterns on digraphs. In
digraphs, a matching is a subset of edges such that no two edges in the set 
share a common starting vertex or a common ending vertex \cite{Bang-Jensen:bg07}. Similarly to the 
Karp--Sipser algorithm, used for constructing of a matching on
undirected graphs, 
in the algorithm for digraphs we choose recursively one directed edge $(i,j)$ 
in which 
vertex $i$ is leaf. This edge is included in the matching and
removed from the graph together with outgoing edges of vertex $j$, if vertex $i$ is in-leaf, or together with incoming edges of vertex $j$, if vertex $i$ is out-leaf. This allows one avoiding 
forbidden configurations in the
matching.

The simplicity of the leaf removal algorithm for directed networks enables us to describe the evolution of the structure of the network during the pruning process by applying rate equations. 
Similar, though significantly more compact, rate equations have been derived for the description of 
the leaf removal algorithm on undirected graphs, while 
constructing minimum vertex covers 
\cite{Weight:whbook2005,weigt}. 
Using these rate equations we re-obtain the size 
of a core in the configuration model of undirected random networks and the number of links in this core.  For the Erd\H{o}s--R\'enyi undirected random graphs, we compare our results for cores with those obtained by the generating function technique \cite{Bauer} and demonstrate the preciseness of the rate equation approach.  
In this work we apply the extended leaf removal algorithm
to the configuration model of directed uncorrelated networks with a given degree distribution and derive rate equations which allow us to find 
the structure of the cores in directed networks. For directed networks with Poisson and power-law degree distributions we compare our results with simulations in Ref.~\cite{control} and observe a complete agreement. 
In- and out-leaves in directed networks can be classified according to the number $m$ of their outgoing and incoming edges, respectively. 
This 
enables us to naturally decompose 
a directed network, layer by layer, removing leaves with higher and higher $m$, to sub-cores; we name them
$m$-cores. 
Note that this decomposition of
the directed networks differs principally from the well-known $k$-core decomposition of networks 
\cite{seidman}.

This paper is organized as follows. 
In the next section, for the sake of clarity, we first apply the rate equation approach for leaf removal \cite{Weight:whbook2005,weigt} to the core extraction problem for undirected uncorrelated networks and demonstrate the preciseness of this approach. 
In Sec.~\ref{directed_sec} we
introduce in detail the extended leaf removal algorithm for directed networks and
derive the rate equations. By using these equations we find the basic characteristics of cores in diverse directed uncorrelated networks 
and explore the $m$-core decomposition of directed networks. 



\section{Undirected networks}
\label{undirected_sec}

\subsection{Rate equations}

Let us first consider an undirected uncorrelated network with $N$ vertices, $L$
edges and a degree distribution $P(q)$, which is described here by the configuration model \cite{Bender:bc78,Bollobas:b80}, i.e., a labelled random graph with a given degree sequence.  
We
apply the leaf removal algorithm 
\cite{Bauer,matching} 
to this network. According to this algorithm, at each step 
we choose at random a leaf and remove it together with its nearest neighbor vertex and all the incident edges to this neighbor. 
The procedure is iterated until no leaves remain in the network.

The structural evolution of the network during the leaf removal process is 
described by so-called rate equations for 
the degree distribution of the remaining network \cite{Weight:whbook2005,weigt}. For the sake of comparison with directed networks, let us remind its meaning (for more detail, see the original derivation in Refs.~\cite{Weight:whbook2005,weigt}). Let $N(q,t)$ be
the average number of vertices with degree $q$ and $N(t)$ be the total
number of remaining vertices at rescaled time $t=T/N$, where
$T$ is the number of algorithmic steps. 
So $\Delta t = 1/N$ as a rescaled time of one iteration.  
The change of the average number of 
vertices with degree $q$ after one step is described by the following expression: 
\begin{eqnarray}
&& 
\!\!\!\!\!\!\!\!
N(q,t+\Delta t)=N(q,t)-\delta_{q,1}
\nonumber
\\[5pt]
&&
\!\!\!\!\!\!\!\!
-\frac{qP(q,t)}{\langle q\rangle_{t}}+\frac{\langle q(q-1)\rangle_{t}}{\langle q\rangle_{t}} \frac{(q+1)P(q+1,t)-qP(q,t)}{\langle q\rangle_{t}}
. 
\label{master}
\end{eqnarray}
The delta function in the right-hand side shows that at each step one leaf is removed. 
In an uncorrelated 
network, the degree
distribution of the end vertices of a uniformly randomly chosen edge is 
$P^{(1)}(q,t)=qP(q)/\langle q\rangle$. So when we remove the
nearest neighbor of the leaf, the number of vertices with degree
$q$, is decreased by $1$ with probability $P^{(1)}(q,t)$. 
After the removal of the leaf and its nearest neighbor with their edges, the number of connections of the second-nearest neighbors of the leaf diminishes. 
So finally the last term describes the change of $N(q)$ due to the decrease by $1$ of the 
degrees 
of the second-nearest neighbors of the removed leaf. 
The number of the second neighbors, in average is equal to the mean degree of the nearest
neighbor vertex except one (connection to the leaf), that is, 
$\sum_{q}(q-1)P^{(1)}(q,t)=\langle q(q-1)\rangle_{t}/\langle q\rangle_{t}$. 
Note that the last term on the right-hand side of Eq.~(\ref{master}) can be derived strictly implementing a single assumption, namely, the absence correlations in the network during the pruning process. 


\begin{figure*}[t]
\begin{center}
\scalebox{0.263}{\includegraphics[angle=0]{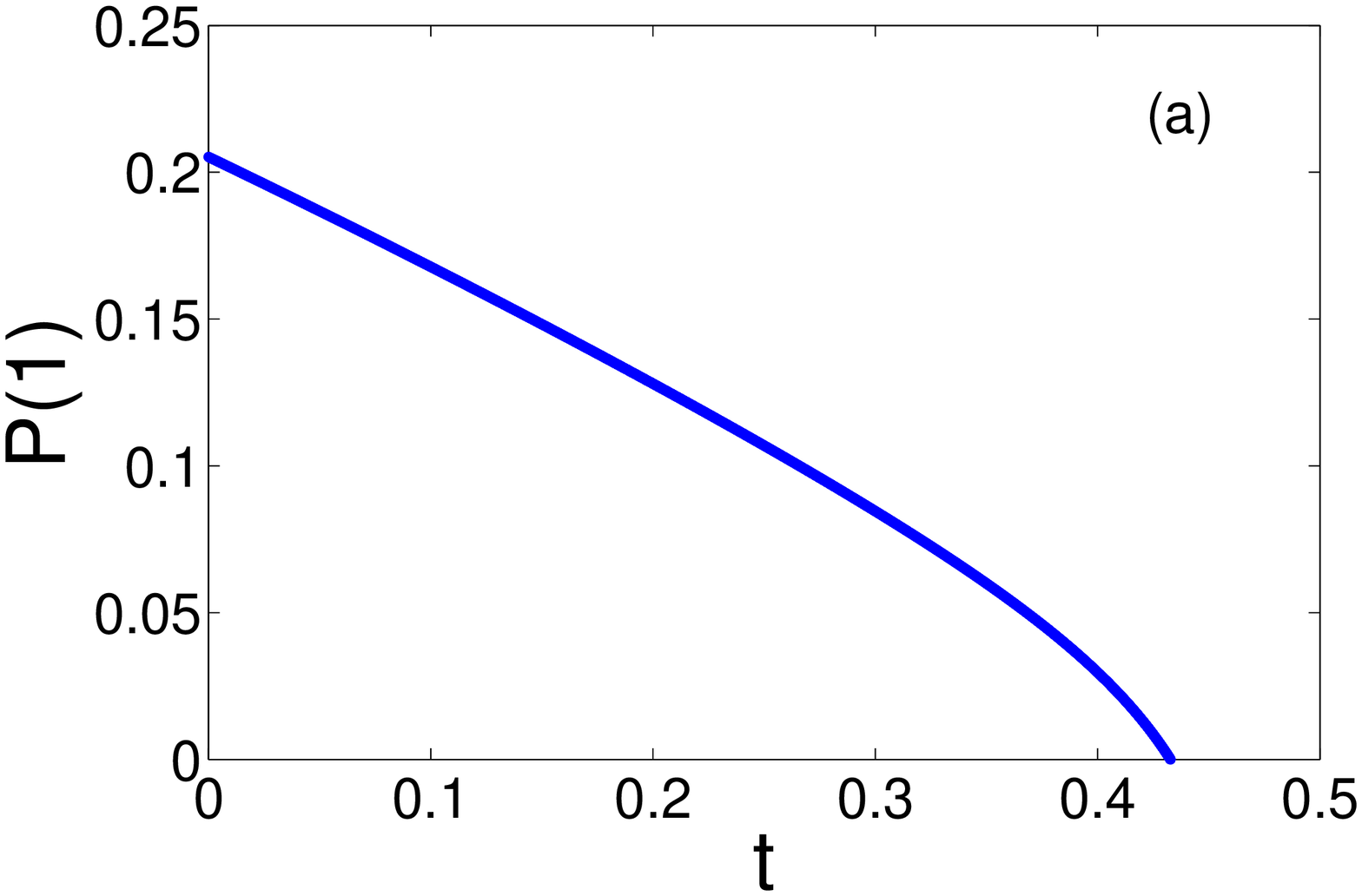}}
\scalebox{0.263}{\includegraphics[angle=0]{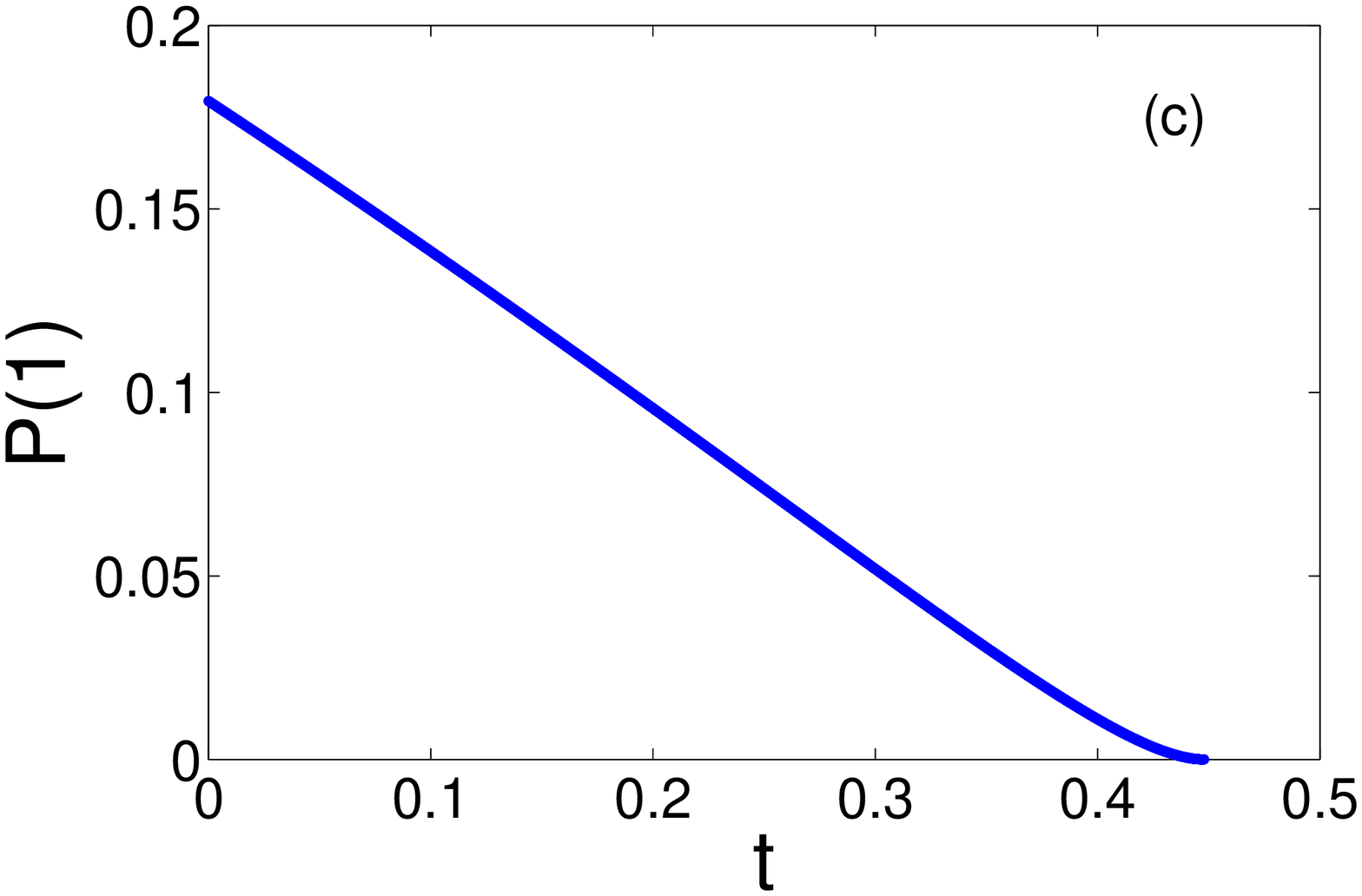}}
\scalebox{0.263}{\includegraphics[angle=0]{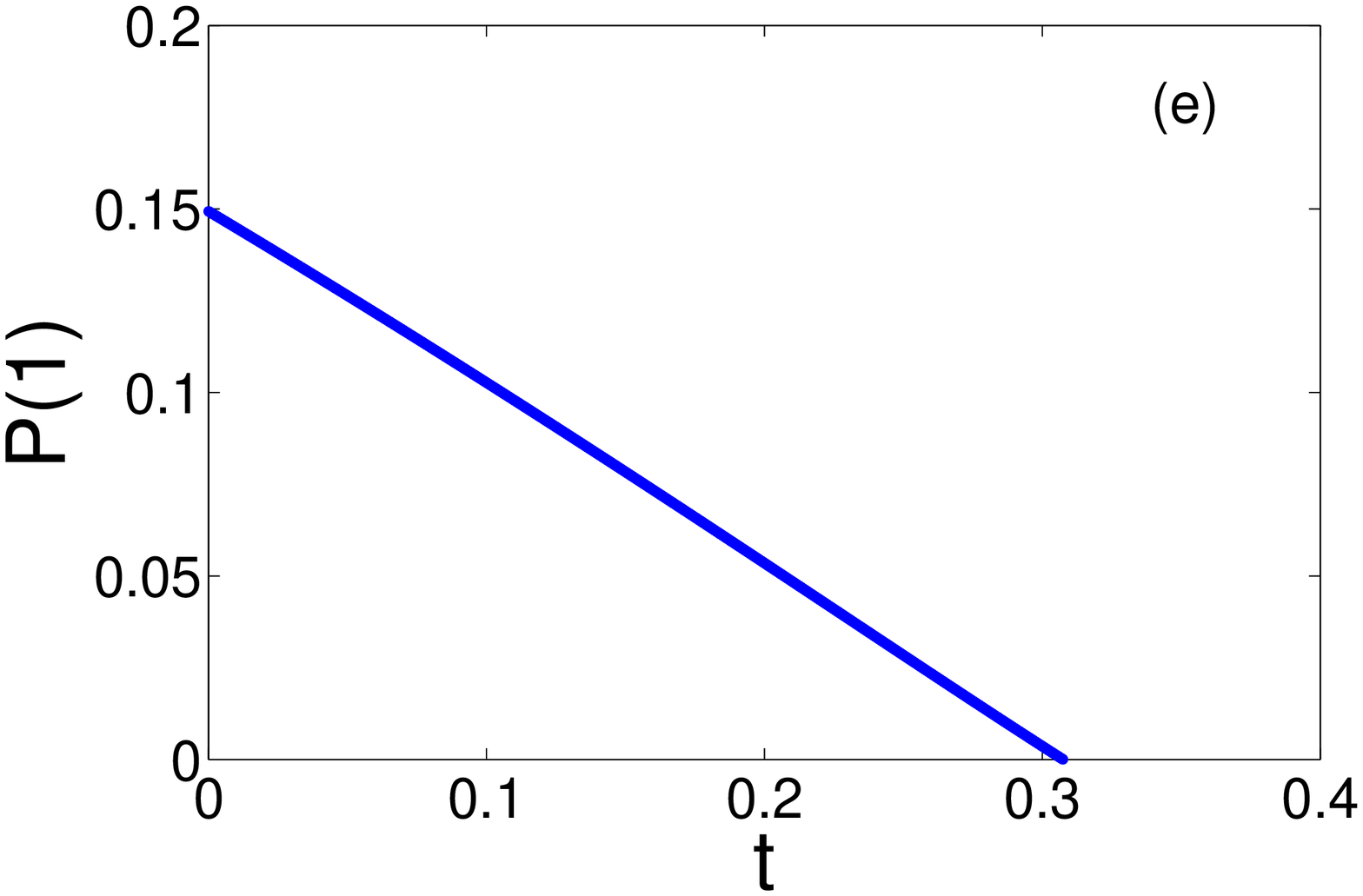}}
\\[-2pt]
\scalebox{0.263}{\includegraphics[angle=0]{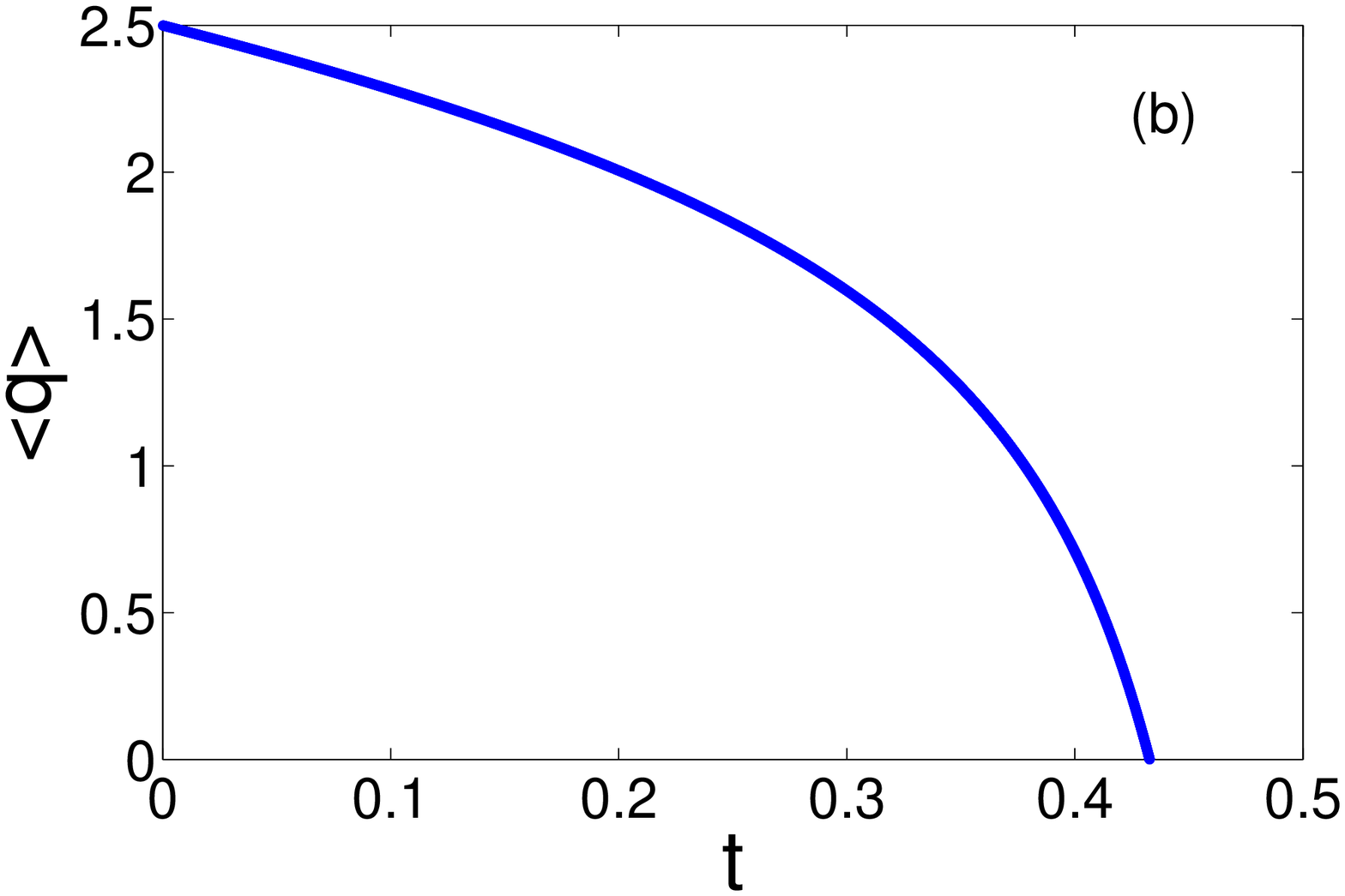}}
\scalebox{0.263}{\includegraphics[angle=0]{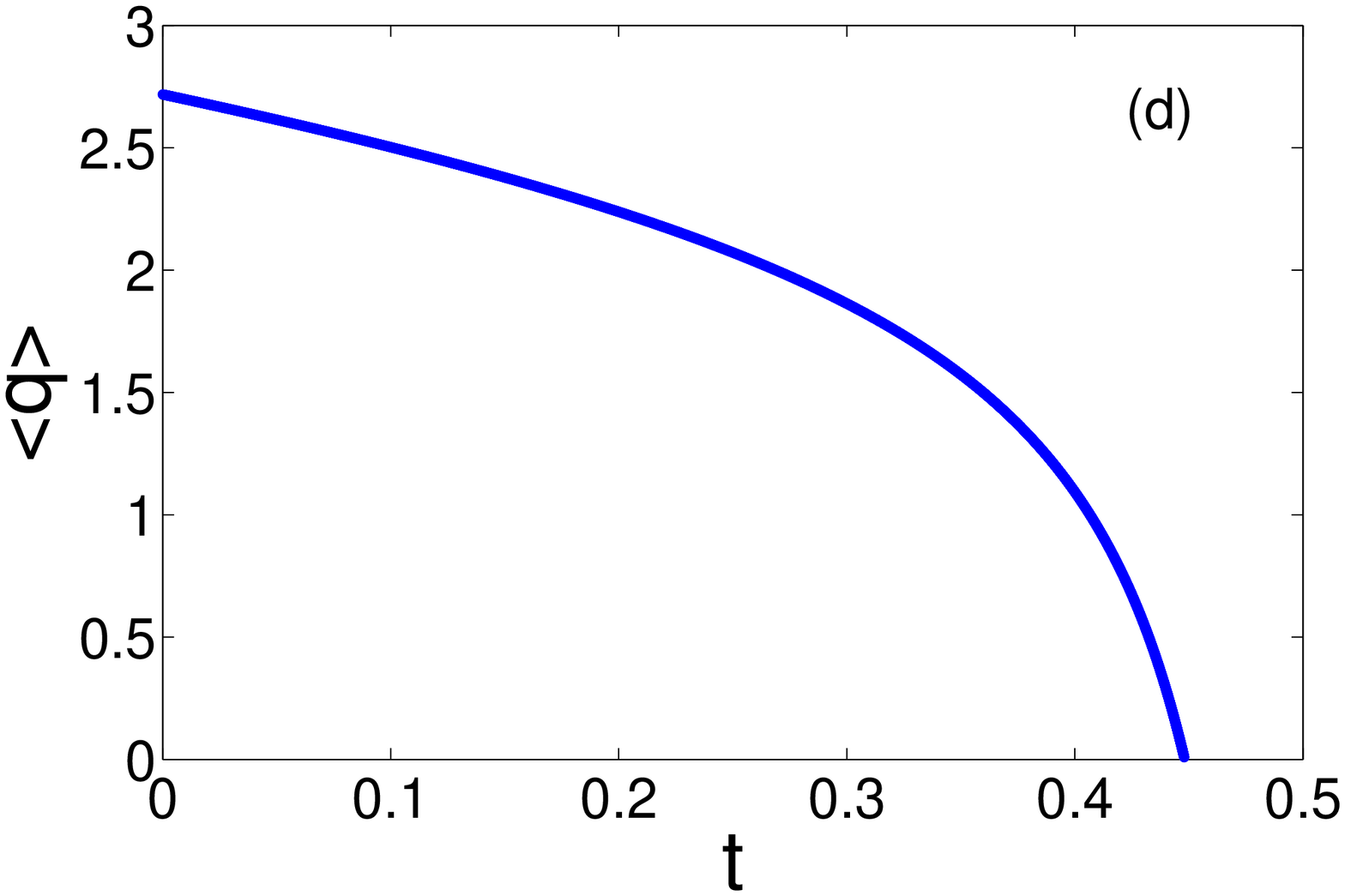}}
\scalebox{0.263}{\includegraphics[angle=0]{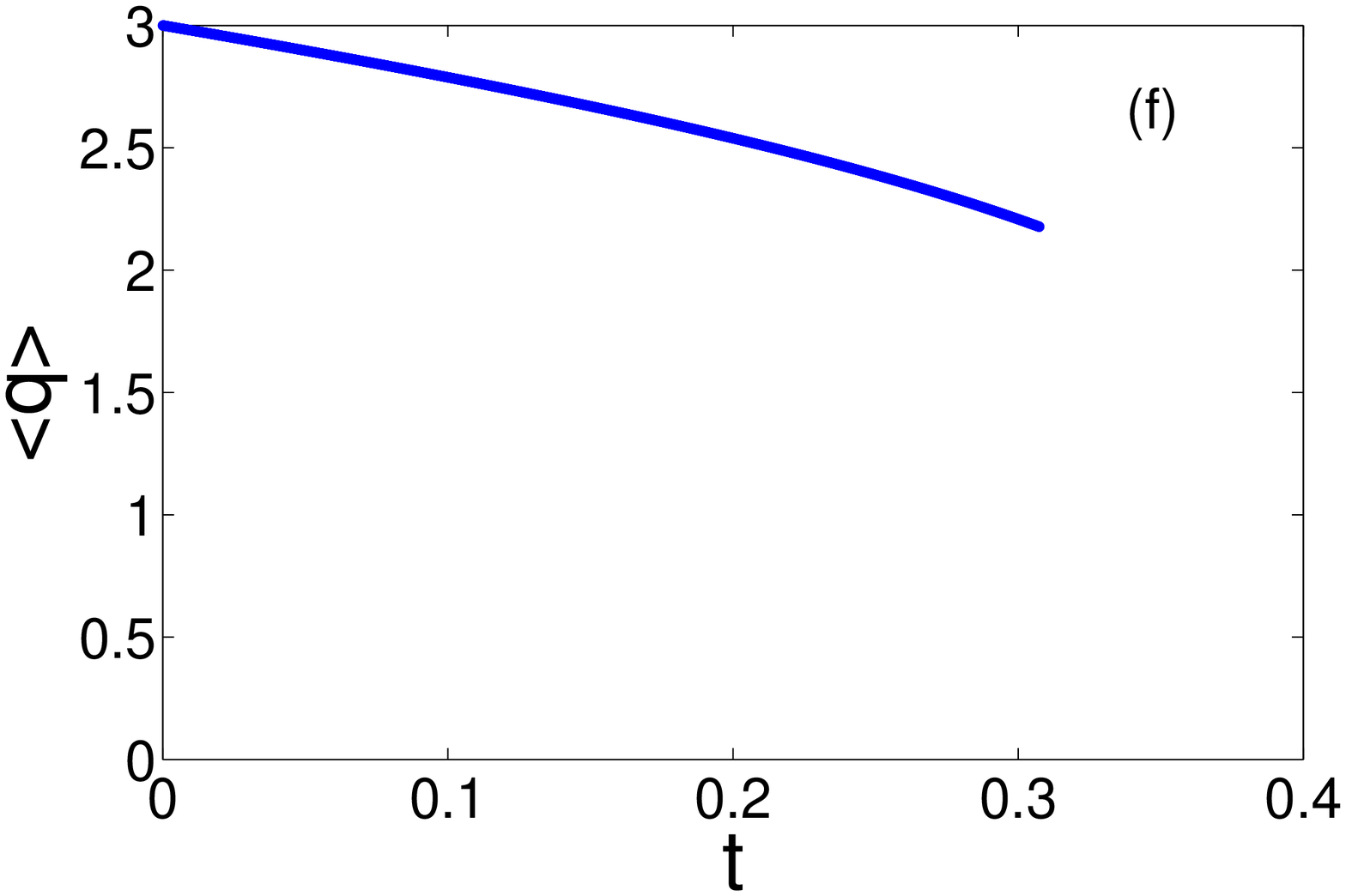}}
\end{center}
\caption{
(Color online) 
Different regimes of the evolution  
of the fraction of leaves $P(1)$ and the corresponding mean
degree generated by the leaf removal algorithm applied to the Erd\H{o}s--R\'enyi undirected graph with various vertex mean degrees $c_0$: (a) and (b) $c_0=2.5$ is below the critical point, (c) and (d) $c_0$ takes the critical value equal $e\approx2.72$,  
(e) and (f) $c_0=3.0$ is above the critical value.  
}
\label{f1}
\end{figure*}


At each step, the total number of the vertices in the network decreases by $2$. The number of removed edges, in average is
equal to the mean degree of the first neighbor of the leaf. 
This results in the 
following 
equations:
\begin{eqnarray}
\frac{\dot{N}(t)}{N}&=&-2 
, 
\label{mastern}
\\[5pt]
\frac{\dot{L}(t)}{N}&=&-\frac{\langle q^{2}\rangle_{t}}{\langle q\rangle_{t}}
.
\label{e3}
\end{eqnarray}
Substituting $N(q,t)=P(q,t)N(t)$ into Eq.~(\ref{master}) gives the following evolution equation for the degree distribution \cite{Weight:whbook2005,weigt}: 
\begin{eqnarray}
&&
(1-2t)\dot{P}(q,t)=2P(q,t)-\delta_{q,1}
\nonumber
\\[5pt]
&&
-\frac{\langle
q^{2}\rangle_{t}}{\langle q\rangle^{2}_{t}}qP(q,t)+\frac{\langle
q(q-1)\rangle_{t}}{\langle q\rangle^{2}_{t}}(q+1)P(q+1,t)
.
\label{masterprob}
\end{eqnarray}
Eq. (\ref{masterprob}) is actually a set of differential equations,
describing the evolution of a network 
during pruning. 
These equations are seemingly similar to the rate equations 
describing the evolution of networks driven by the preferential attachment mechanism \cite{growth1, growth2}. 
The key difference however is that Eq.~(\ref{masterprob}) is nonlinear due to the moments of the degree distribution $P(q,t)$ which enter the right-hand side of the equation. This nonlinearity makes the solution of the equation particularly difficult.   
Notably, while deriving, it was assumed that the network remains uncorrelated during this pruning process. This is actually a strong assumption, which was never specially proved. For example, in a well-studied problem of connected components in uncorrelated networks, the giant component generally have degree--degree correlations \cite{Bialas:bo08}. This is why it is interesting to compare results for cores obtained by using the rate equations with those found by other methods. 

The set of equations (\ref{masterprob}) was treated analytically only while studying vertex covers for the Erd\H{o}s--R\'enyi random graphs \cite{weigt}. The analysis was based on the following ansatz. It was assumed that the degree distribution stays Poissonian for $q>1$ during the whole leaf removal process (though mean degree varies). In the present paper we use a different approach, without making this assumption. We directly solve numerically the set of these equations ($0 \leq q \leq q_{\text{max}}$)
and obtain the structure of the core 
analyzing the evolution of the degree distribution during the pruning process. Note that we consider infinite networks which may have vertices with degrees approaching infinity. In this situation we have to truncate the set of evolution equations. 


\begin{figure}[t]
\begin{center}
\scalebox{0.271}{\includegraphics[angle=0]{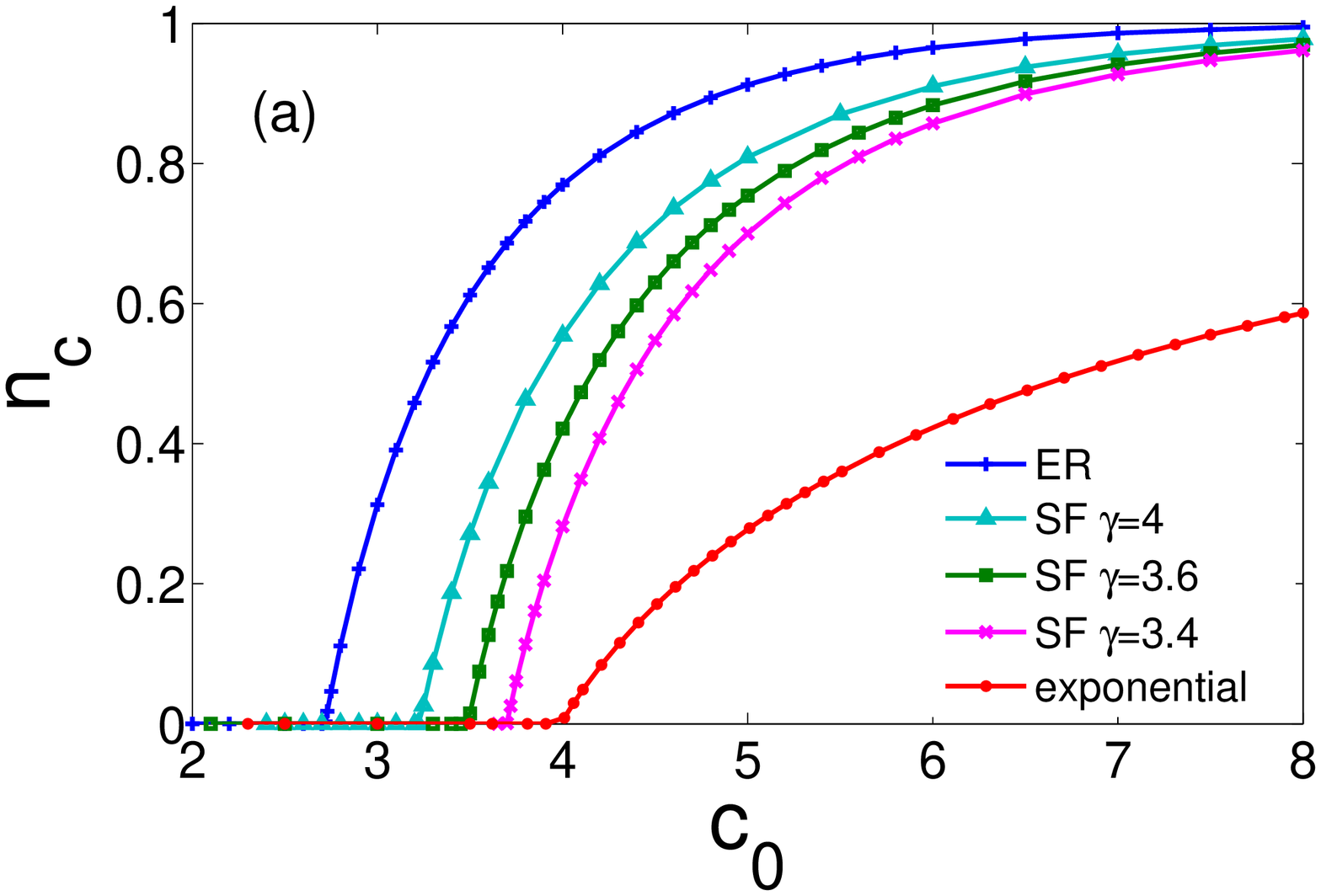}}
\\[-2pt]
\scalebox{0.271}{\includegraphics[angle=0]{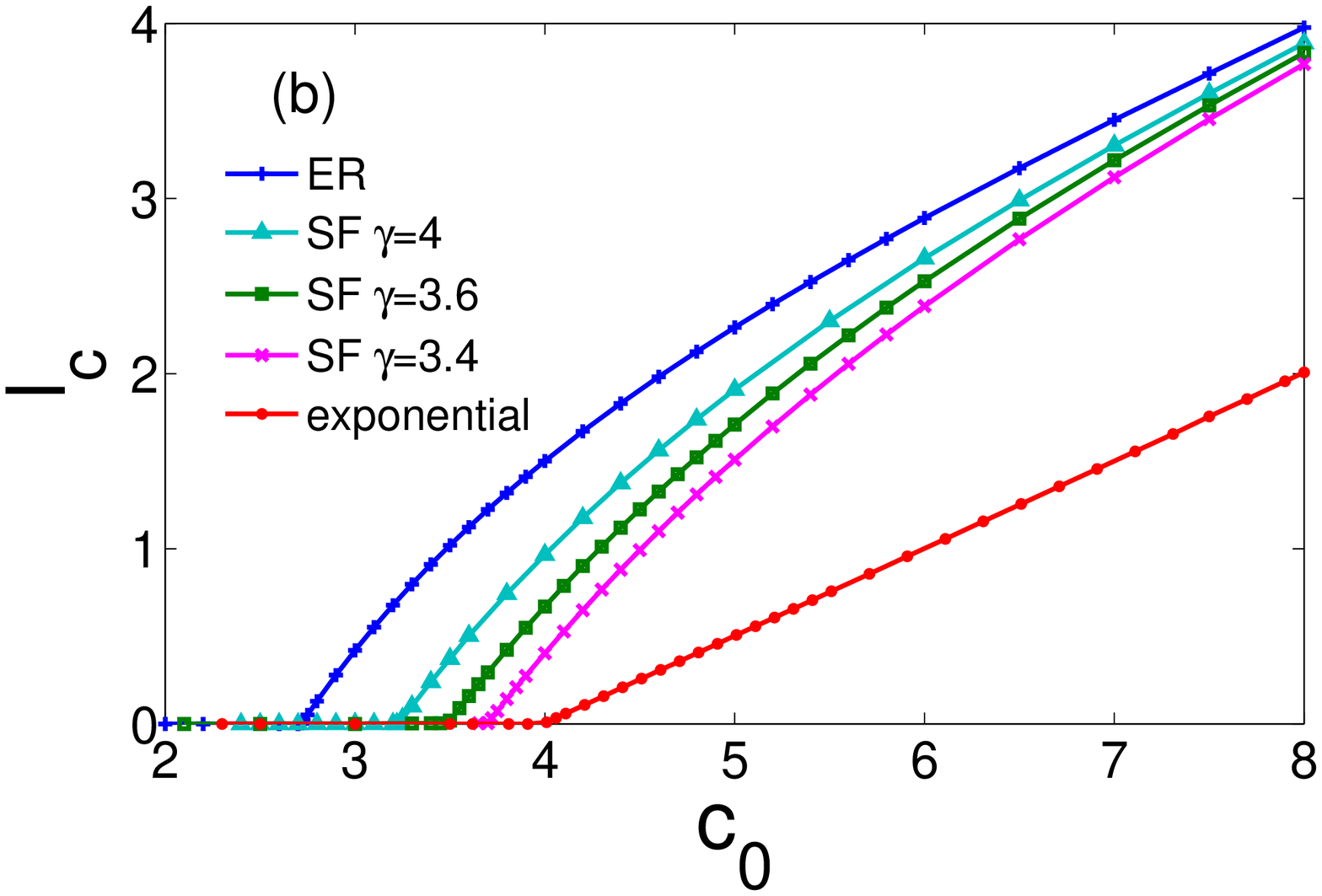}}
\\[-2pt]
\scalebox{0.271}{\includegraphics[angle=0]{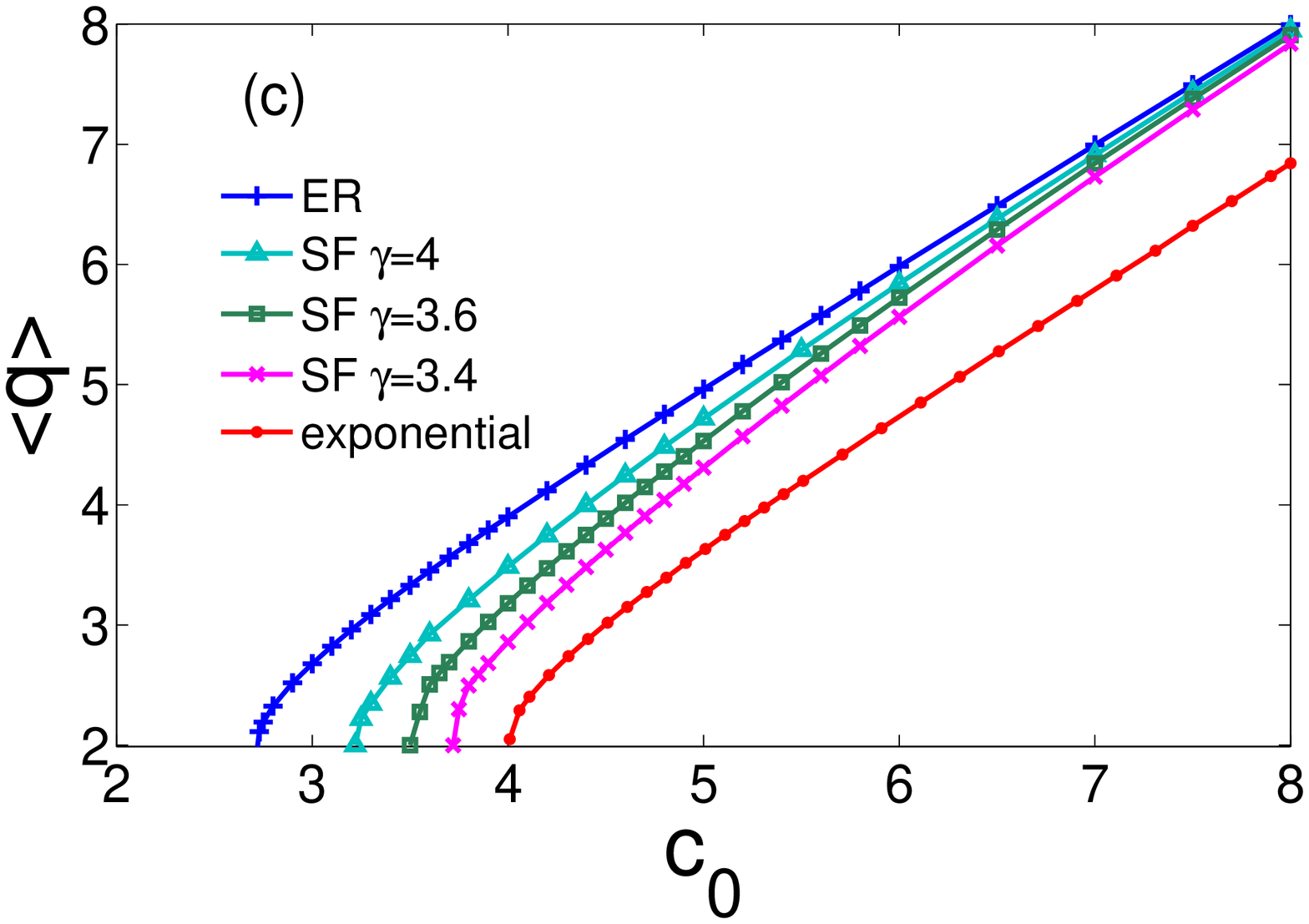}}
\\[-2pt]
\scalebox{0.271}{\includegraphics[angle=0]{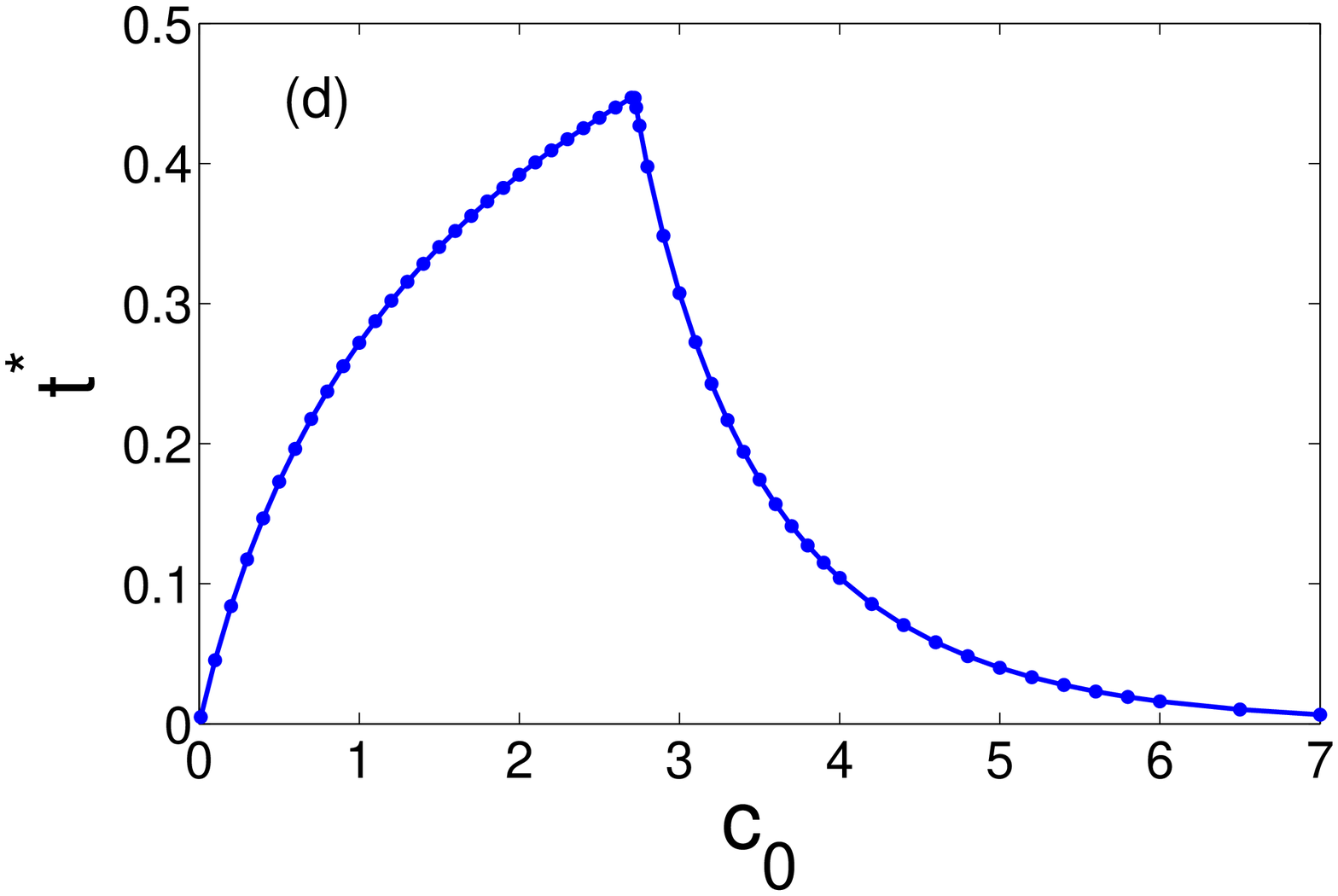}}
\end{center}
\caption{ 
(Color online) 
(a) The relative core size $n_c=N_c/N$, (b) the normalized number of edges $l_c=L_c/N$ in a 
core, and (c) the mean vertex degree $\langle q \rangle=2L_c/N_c$ of a core for different undirected networks 
vs the 
vertex mean
degree $c_0$ of an original network. 
The networks are the Erd\H{o}s--R\'enyi random graph (ER), the scale-free networks (SF) with the degree distribution exponents $\gamma=4, 3.6, 3.4$, and the network with an exponentially decaying degree distribution.   
(d) Normalized time $t^*$, at which the pruning process finishes 
vs the 
vertex mean
degree $c_0$ of an original network for the Erd\H{o}s--R\'enyi 
graphs.
}
\label{f2}
\end{figure}



\begin{figure}[t]
\begin{center}
\scalebox{0.273}{\includegraphics[angle=0]{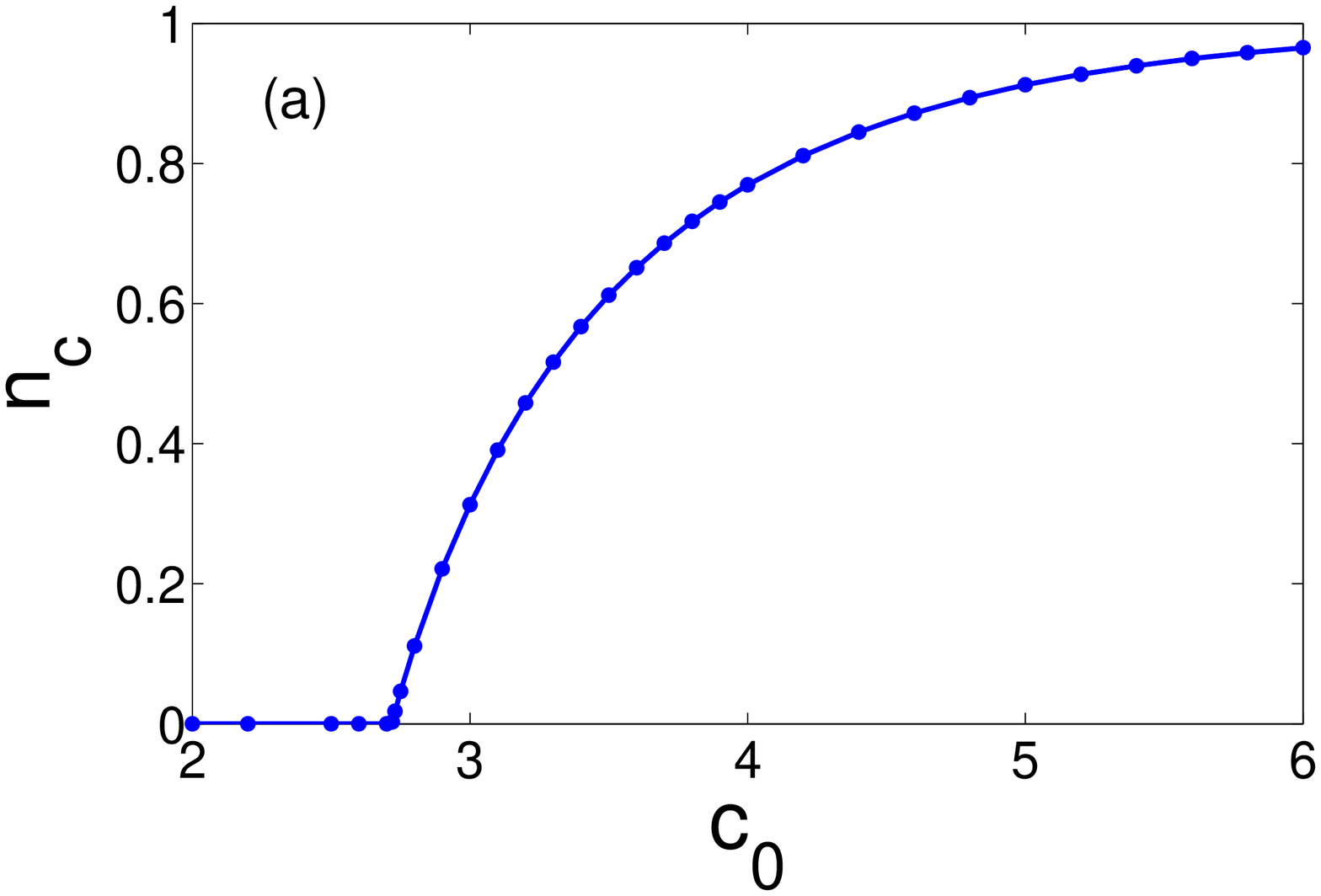}}
\scalebox{0.273}{\includegraphics[angle=0]{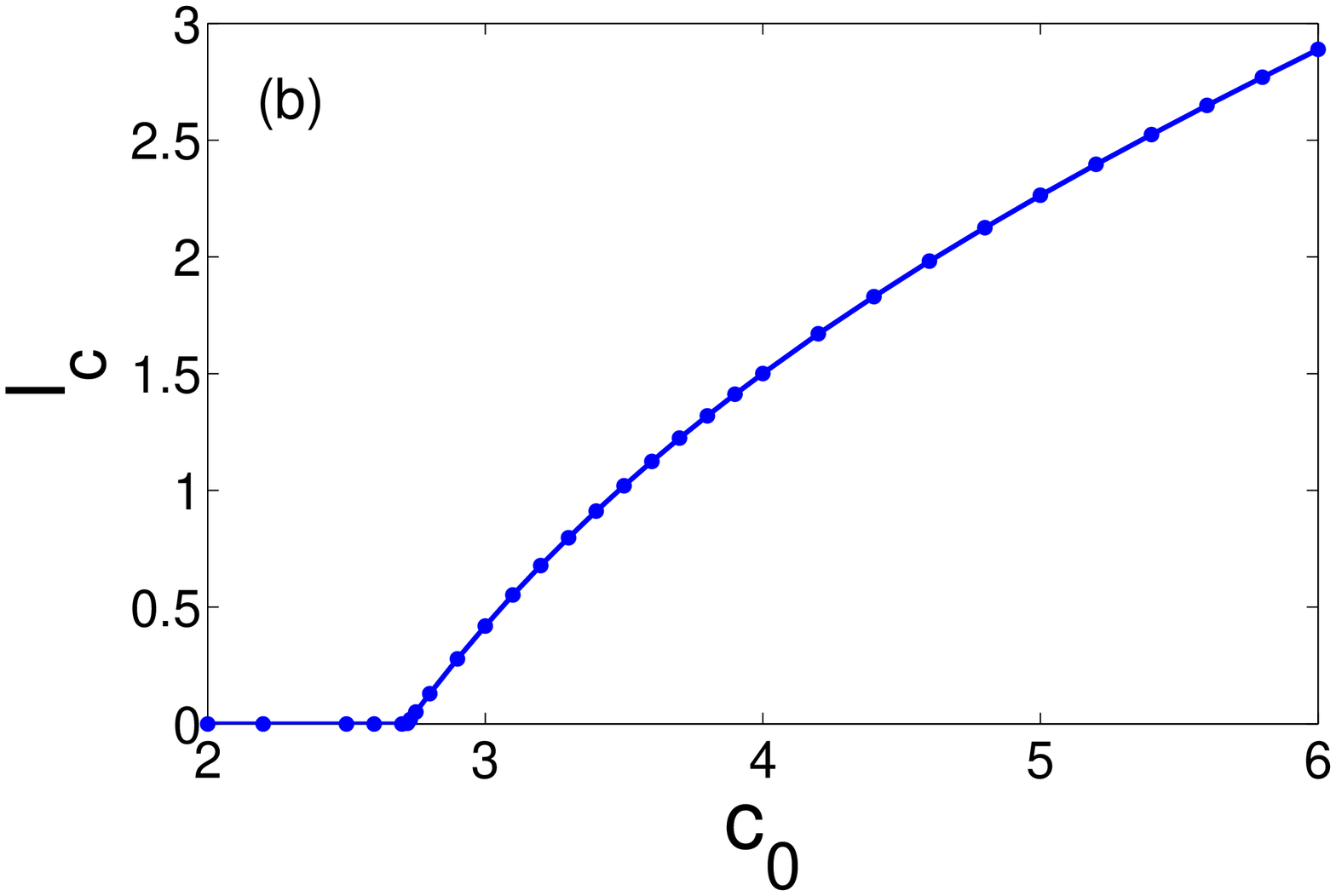}}
\end{center}
\caption{
(Color online) 
Comparison of our results (dots) for the Erd\H{o}s--R\'enyi undirected graphs, namely, the relative core size $n_c(c_0)$ (a) and the normalized number of edges $l_c(c_0)$ in the core (b), with the corresponding results 
obtained by the generating function technique (solid lines) \cite{Bauer}.  
}
\label{f3}
\end{figure}



\subsection{Cores in undirected networks}

We apply the leaf removal algorithm to an undirected uncorrelated network with degree distribution $P(q,t=0)$ and a vertex mean degree equal to $c_0$, as initial conditions. Algorithm is iterated until no vertices of degree $1$ remain. The degree distribution of the network is obtained at each time step by solving Eq. (\ref{masterprob}).
We focus on the evolution of 
the relative number of leaves, 
$P(1,t)$. 
Changing the initial mean degree $c_0$,  
we
observe two different behaviors for $P(1,t)$ 
separated by a critical point, see Fig.~\ref{f1}. 
Below the critical point, $P(1,t)$
becomes zero exactly in the moment in which the vertex mean degree in the remaining networks reaches zero. 
This means that the algorithm finishes when the network disappears. 
Contrastingly, above the critical point, $P(1,t)$ becomes zero at a time $t^*$ at which the mean degree is still nonzero. That is, at this moment, there still remains 
a subgraph 
with vertices of degrees greater than $1$, 
i.e., the core. The number of remaining vertices except isolated
vertices 
at time $t^*$, is the size
of the core:
\begin{eqnarray}
\!\!\!\!\! 
N_{c}= N(t^{*})-N(0,t^{*})&=& N(t^{*})-N(t^{*})P(0,t^{*})
\nonumber
\\[5pt]
\!\!\!\!\! 
&=&N(1-2t^{*})[1-P(0,t^{*})] 
.
\label{sizecore}
\end{eqnarray}
The number of edges in the core is obtained from Eq.~(\ref{e3}) into which we substitute the solution $P(q,t)$ of Eq.~(\ref{masterprob}), namely,   
\begin{equation}
L_{c}=L(t^{*})
.
\label{edgecore}
\end{equation}

To demonstrate this approach, we apply it to a few undirected uncorrelated networks. First, consider the 
Erd\H{o}s--R\'enyi random graphs, which have a Poisson degree distribution,  
$P(q) = e^{-c_{0}}c_{0}^{q}/q!$, in which $c_{0}= 2 L/N$ is
the mean vertex degree for the graph. 
The rapid decay of the Poisson distribution allows us to reduce greatly the set of the evolution equations which we solve numerically. For the mean degrees $c_0$ in the range between $1$ and $8$, it is sufficient to solve the set of the first $21$ equations, $q_{\text{max}}=20$. 
Figure~\ref{f1} demonstrates three different regimes of the leaf removal process. Note that at the critical point, $c_0\approx 2.72\approx e$, the curve $P(1,t)$ is tangent to the $t$ axis at $t^*$, see Fig.~1(c). 
In Fig.~\ref{f2}(a,b) we show the relative size of a core, $n_{c}= N_{c}/N$ and the 
normalized number of edges in a core, $l_{c}= L_{c}/N$ for several uncorrelated networks including 
the Erd\H{o}s--R\'enyi undirected graphs. The points on the plots 
are obtained in the following 
way: for each considered value of initial mean degree of the network $c_{0}$,
we find the time $t^{*}$ at which the leaf removal process is completed and the evolution of the distribution $P(q,t \leq t^*)$ and so of its moments, and 
then the size and the number of edges are
calculated from Eqs.~(\ref{e3}), (\ref{sizecore}), and (\ref{edgecore}). For the Erd\H{o}s--R\'enyi graph, our
approximate value for the critical point is $c_{0} \approx 2.72$, which 
agrees 
with 
Refs.~\cite{Bauer,corenew}. 
The mean vertex degree $\langle q \rangle = 2L_c/N_c$ of a core approaches $2$ at the critical point revealing a long-loop structure, see Fig.~\ref{f2}(c), also indicated in Ref.~\cite{Bauer}. 
For the Erd\H{o}s--R\'enyi graphs, the normalized time $t^*$ at which the pruning process finishes 
vs the original mean vertex degree $c_0$ is shown in Fig.~\ref{f2}(d). Clearly, at small $c_0$, almost all edges have no joint vertices, so the process will finish after $T^*=L$ steps, so $t^* \cong L/N =c_0/2$. There is a sharp change of the dependence $t^*(c_0)$ at the critical point. 
Note that below the critical point, $t^*$ coincides with the normalized size of the maximum matching $M_{\text{max}}/N$. 
In contrast, above the critical point, i.e., in the presence of the core, $t^*$ provides the normalized size of only the part of the maximum matching sitting outside of the core.  
For comparison, 
the size of the maximum matching of an entire Erd\H{o}s--R\'enyi graph above the critical point is rather close to the maximum matching size value at the critical point. This follows from the result of the Karp--Sipser algorithm applied to the Erd\H{o}s--R\'enyi graphs 
\cite{matching}, which, above the critical point, gives an estimate of the size of the maximum matching.  
Furthermore, the comparison of the dependencies of the size and the number of edges in the core on the vertex mean degree $c_0$ found by the numerical solution of the rate equations and those obtained by using generating functions reveals a complete agreement between these two approaches, see Fig.~\ref{f3}(a) and (b). This suggests that the evolution equations, which were used, are actually asymptotically exact, similarly to the approach of Ref.~\cite{Bauer}.  

We also considered  
uncorrelated networks with an exponential degree distribution
$P(q)=(1-e^{-1/\kappa})e^{-q/\kappa}$, 
where 
$c_{0}=e^{-1/\kappa}/(1-e^{1/\kappa})$. 
The critical point is at $c_{0} \approx 4.0$ as is shown in Fig.~\ref{f2}(a,b). 
In the case of scale-free networks $P(q)\propto q^{-\gamma}$, $n_c$ and $l_c$ depend on both the $\gamma$ exponent and the mean degree $c_0$. 
We consider asymptotically scale-free networks with the degree distribution 
$P(q) =[\frac{c_0(\gamma-2)}{2(\gamma-1)}]^{\gamma-1}\Gamma(q-\gamma+1,\frac{c_0(\gamma-2)}{2(\gamma-1)})/\Gamma(q+1) \cong q^{-\gamma}$,
where $\Gamma(s)$ is the gamma function and $\Gamma(s,x)$ is the
upper incomplete gamma function \cite{static}. 
This degree distribution is typical in the static model \cite{Goh:gkk01}, which is extensively used in simulations. This is why this form will allow for comparison with results obtained in numerical simulations.  
Here we only explore the region $\gamma >3$. 
To obtain the dependencies $n_c(c_0)$ and $l_c(c_0)$ for scale-free networks shown in Fig.~\ref{f2}(a,b) we 
solve Eq. (\ref{masterprob}) 
with 
the cutoff $q_{\text{max}}=100$, ignoring the higher degrees.  
The value of cutoff $q_{\text{max}}$, which we select, depends on the values of the exponent $\gamma$ and the mean degree $c_0$. 
To obtain precise results 
we should 
set sufficiently large cutoffs $q_{\text{max}}$. 
To choose a relevant cutoff value $q_{\text{max}}$ at given $\gamma$ and $c_0$, we have to repeat   
calculations with a few $q_{\text{max}}$ values and to analyse convergence of the results with growing $q_{\text{max}}$.


\section{Directed networks}
\label{directed_sec}

\begin{figure}[t]
\begin{center}
\scalebox{1.05}{\includegraphics[angle=0]{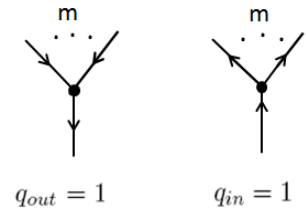}}
\end{center}
\caption{
Various out- and in-leaves 
in directed networks. 
For out- and in-leaves, the number $m$ of, respectively, incoming and outgoing edges is arbitrary, $m=0,1,2,...$. In the special case of $q_{in}=q_{out}=1$, the vertex is simultaneously an in- and out-leaf. 
}
\label{f4}
\end{figure}

In directed networks,
each vertex is characterized by its in-degree 
$q_{in}$ and 
out-degree $q_{out}$, and the joint in,out-degree distribution of a network is 
$P(q_{in},q_{out})$. 
In the 
directed networks, leaves are 
defined as vertices with
$q_{in}=1$ or $q_{out}=1$, see Fig.~\ref{f4}. In general, removing of
all leaves leads to 
a giant core 
above a critical point. 
Importantly, 
the leaf removal algorithm for directed networks should
be 
formulated in such a way that this algorithm will be  
applicable for
combinatorial optimization problems on directed graphs, like the
matching problem. 


\subsection{Leaf removal algorithm: rate equations}

\begin{figure}[t]
\begin{center}
\scalebox{0.9}{\includegraphics[angle=0]{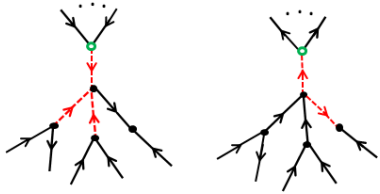}}
\end{center}
\caption{
(Color online) 
Leaf removal process on directed networks. The 
open dots show 
out- (see left) and in- (see right) leaves. 
Once a leaf is selected, the dashed edges are removed, and the leaf becomes a normal vertex. 
Note that only the dashed edge attached to a leaf is added to the matching. 
}
\label{f5}
\end{figure}

Let us generalize the leaf removal algorithm to directed networks. 
By definition, the leaves here are 
vertices with $q_{in}=1$ 
and an arbitrary number (including zero) of outgoing edges 
(in-leaf) or with $q_{out}=1$ and an arbitrary number (including zero) of incoming edges 
(out-leaf). 
In the particular case $q_{in}=q_{out}=1$, the vertex is simultaneously an in- and out-leaf, but this 
coincidence does not affect the leaf removal process. 
The
extended leaf removal algorithm is defined in the following way. At 
each step, one out-leaf or one in-leaf is uniformly randomly
chosen with probability $1/2$. 
The single outgoing edge or incoming of this out- or, respectively, in-leaf is removed, so that this vertex becomes a ``normal'' (not a leaf) vertex  
(notice that
we 
do not delete this vertex  
because it may still have connections with other
its neighbors). 
Next, we focus on the neighbor vertex which was connected with the leaf by that removed edge. 
If the leaf was an out-leaf, we also remove all its neighbor's incoming edges; if that leaf was an in-leaf, we remove all its neighbor's outgoing edges, since they are not allowed to enter in the corresponding matching pattern. Figure~\ref{f5} explains this process. 
This procedure is
iterated until no 
leaves are remained. 
Importantly, in contrast to undirected networks, during this process, the number of vertices in the network is conserved. The leaves, being removed, are transformed into normal vertices. Interestingly, in this procedure, a vertex may change its status several times. For example, the following evolution of a vertex is possible. In the original network the vertex is normal, then it becomes an in-leaf, afterwards it is transformed again into a normal vertex, later it becomes an out-leaf, and finally it is an isolated vertex. 

Another marked difference from undirected networks is that in directed networks leaves can have various numbers of edges, see Fig.~\ref{f4}. 
We propose to distinguish and classify leaves in directed networks by the number of their connections. In our leave removal algorithm we recursively remove leaves having $q_{out}=1, q_{in} \leq m$ or $q_{in}=1, q_{out} \leq m$ (we call them $m$-leaves), which results in the specific subgraph which we name $m$-{\em core}. Here $m$ is a given non-negative integer. If $m$ equals the highest vertex in,out-degree in the graph, i.e., $max(q_{in},q_{out})$, the algorithm surely 
removes all possible leaves from a network, which finally results in the main (or central) core or, as we call it here, simply, the core. 
In fact, this can happen already at some smaller value of $m$, which depends on the complete structure of a network.

Let us derive a master equation, describing
the evolution of the structure of an uncorrelated directed network during this algorithm.
Let $N(q_{in}, q_{out},t)$ be the average number of vertices with
in-degree $q_{in}$ and out-degree $q_{out}$ at time $t$. As a
consequence of the removal of edges, $N(q_{in}, q_{out},t)$  
is changed after each
iteration. 
The dynamics can be decomposed into three steps. 


(1) Choosing a leaf. 
We choose uniformly at random vertices with degrees $q_{out}=1, q_{in} \leq m$ (out-leaves) or $q_{in}=1, q_{out} \leq m$ (in-leaves). At each step we choose one leaf. Suppose that we selected  an out-leaf (this choice is made with probability $1/2$). The probability that the in-degree of this leaf turns out to be $q_{in}$, is equal to 
$\theta(m-q_{in})P(q_{in},1)/[\sum_{q_{in} \leq m}P(q_{in},1)]$.  
Here $\theta(i)$ is defined for integers: $\theta(i{\geq}0)=1$ and $\theta(i<0)=0$. 
We must remove the outgoing edge of this leaf, and so with
this probability the number of vertices with $q_{out}=1, q_{in}\leq m$
decreases by $1$ while the number of vertices with $q_{out}=0, q_{in}\leq m$
increases by $1$. 
In the expression for the difference $N(q_{in},q_{out},t+\Delta t)-N(q_{in},q_{out},t)$ [see Eq.~(\ref{masterdir}) below], this gives the first two terms on the right-hand side. Similar arguments are valid if the selected leaf is in-leaf. Two corresponding terms can be seen on the right-hand side of Eq.~(\ref{masterdir}). 


(2) Changing the degree of that leaf's nearest neighbor which is connected to the
leaf by a single incoming or outgoing edge. The probability
that this outgoing edge of the out-leaf at its second end has a vertex of degree $q_{in}, q_{out}$
is 
$ P_{out}^{(1)}(q_{in}, q_{out})= q_{in}P(q_{in},q_{out})/\langle q_{in} \rangle$. 
Therefore with this probability the
number $N(q_{in}, q_{out})$ is decreased by one, which gives the third term on the right-hand side of Eq.~(\ref{masterdir}). 
Recall that we
remove all incoming edges to this
vertex, and so its in-degree  
changes to $q_{in}=0$. 
Consequently, the number of vertices with $q_{in}=0$ increases by
$\delta_{q_{in},0}\sum_{q_{in}}q_{in}P(q_{in},q_{out})/\langle q_{in} \rangle$, which is the fourth term on the right-hand side of Eq.~(\ref{masterdir}). 
A similar argument is valid when we select an in-leaf.

(3) Changing the degree of leaf's second neighbors. When we select an
out-leaf and remove 
its outgoing edge, we also remove all incoming connections 
from the leaf's neighbor on this edge, see Fig.~\ref{f5}.  
Consequently, the out-degrees of the corresponding leaf's second neighbors decrease by $1$. 
The number of these incoming edges except the edge
attached to the leaf, in average is equal to the mean in-degree of the
nearest neighbor of the leaf minus $1$, i.e., $\sum
_{q_{in}, q_{out}} (q_{in}-1)P_{in}^{(1)}(q_{in}, q_{out})$. 
This leads to the fifth term on the right-hand side of Eq.~(\ref{masterdir}).
Similar arguments are valid when we select an in-leaf. 

Applying all these arguments we finally obtain the evolution equation for the degree distribution $P(q_{in}, q_{out},t)$:
\begin{widetext}
\begin{eqnarray}
&&
 N(q_{in},q_{out},t+\Delta t)-N(q_{in},q_{out},t) = 
\dot{P}(q_{in}, q_{out},t) 
\nonumber 
\\[5pt] 
&&
= \frac{1}{2} \Bigg(- \frac{\delta _{q_{out},1}\theta(m-q_{in})P(q_{in},1,t)}{\sum _{q_{in}\leq m}
P(q_{in},1,t)}+\frac{\delta _{q_{out}, 0}\theta(m-q_{in})P(q_{in},0,t)}{\sum _{q_{in}\leq m}
P(q_{in},0,t)}
 - \frac{q_{in}P(q_{in}, q_{out},t)}{\langle q_{in} \rangle_{t}}
\nonumber
\\[5pt] 
&&
+\frac{\delta_{q_{in},0}\sum _{q_{in}}q_{in}P(q_{in},q_{out},t)}{\langle q_{in} \rangle_{t}}
 +\frac{\langle q_{in}(q_{in}-1)\rangle_t}{\langle q_{in} \rangle_{t}}\frac{(q_{out}+1)P(q_{in},q_{out}+1,t)-q_{out}P(q_{in},q_{out},t)}{\langle q_{out} \rangle_{t}}\Bigg)
\nonumber
\\[5pt]
&&
+ \frac{1}{2} \Bigg(- \frac{\delta _{q_{in},1}\theta(m-q_{out})P(1,q_{out},t)}{\sum _{q_{out}\leq m}
P(1,q_{out},t)}+\frac{\delta _{q_{in}, 0} \theta(m-q_{out})P(0, q_{out},t)}{\sum _{q_{out}\leq m}
P(0, q_{out},t)}
 - \frac{q_{out}P(q_{in}, q_{out},t)}{\langle q_{out} \rangle_{t}}
\nonumber
\\[5pt]
&&
+\frac{\delta_{q_{out},0}\sum _{q_{out}}q_{out}P(q_{in},q_{out},t)}{\langle q_{out} \rangle_{t}}
 +\frac{\langle q_{out}(q_{out}-1)\rangle_t}{\langle q_{out} \rangle_{t}}\frac{(q_{in}+1)P(q_{in}+1,q_{out},t)-q_{in}P(q_{in},q_{out},t)}{\langle q_{in} \rangle_{t}}\Bigg)
.
\label{masterdir}
\end{eqnarray}
\end{widetext}
We used the fact that, no vertices are removed in this process in contrast to undirected networks, i.e.,  
$N(t+\Delta t)= N(t)= N$. The time step is $\Delta t = 1/N$, and in the large network limit we passed from the discrete difference to the time derivative of the degree distribution $P(q_{in}, q_{out},t)= N(q_{in}, q_{out},t)/N$. 
Equation~(\ref{masterdir}) is actually an infinite set of nonlinear
differential equations, even more difficult to solve than  
Eq. (\ref{masterprob}) for undirected
networks. 

It turns out possible to express the time derivative of total number of edges of the network in terms of the moments of the degree distribution.  
If we choose an 
out-leaf, we remove all incoming edges of the leaf's nearest
neighbor (this neighbor is at the second end of the outgoing edge of the leaf). The number of these edges in average is equal to the mean
in-degree of the leaf's nearest neighbor vertex. The probability
that we reach a vertex of degree $q_{in}, q_{out}$, following an
outgoing edge of the leaf is 
$ P_{out}^{(1)}(q_{in}, q_{out}) = q_{in}P(q_{in}, q_{out})/\langle q_{in} \rangle$. 
Therefore,
the mean in-degree of the leaf's neighbor is 
$ \sum _{q_{in},q_{out}}q_{in}P_{out}^{(1)}(q_{in},q_{out}) = 
\langle q_{in}^2\rangle_t/\langle q_{in}\rangle_t$. 
If 
an in-degree leaf is chosen, the number of edges which are removed at one
step in average is equal to the mean out-degree of the leaf's nearest
neighbor, which is, similarly,  
$\langle q_{out}^{2}\rangle_t/\langle q_{out}\rangle_t$. 
Choosing
out-degree or in-degree leaves with equal probability ($1/2$) leads to the
following result: 
\begin{eqnarray}
\frac{\dot{L}(t)}{N}=-\frac{1}{2}\Bigg(\frac{\langle q_{in}^{2}\rangle_{t}}{\langle q_{in}\rangle_{t}}+\frac{\langle q_{out}^{2}\rangle_{t}}{\langle q_{out}\rangle_{t}}\Bigg)
.
\label{masterdiredge}
\end{eqnarray}
Our numerical results for cores in directed networks are based on Eqs.~(\ref{masterdir}) and (\ref{masterdiredge}). One may suggest that they are asymptotically exact, similarly to Eqs.~(\ref{mastern})--(\ref{masterprob}) for undirected uncorrelated networks which we checked, see Figs.~\ref{f3}(a,b).


\begin{figure}[t]
\begin{center}
\scalebox{0.269}{\includegraphics[angle=0]{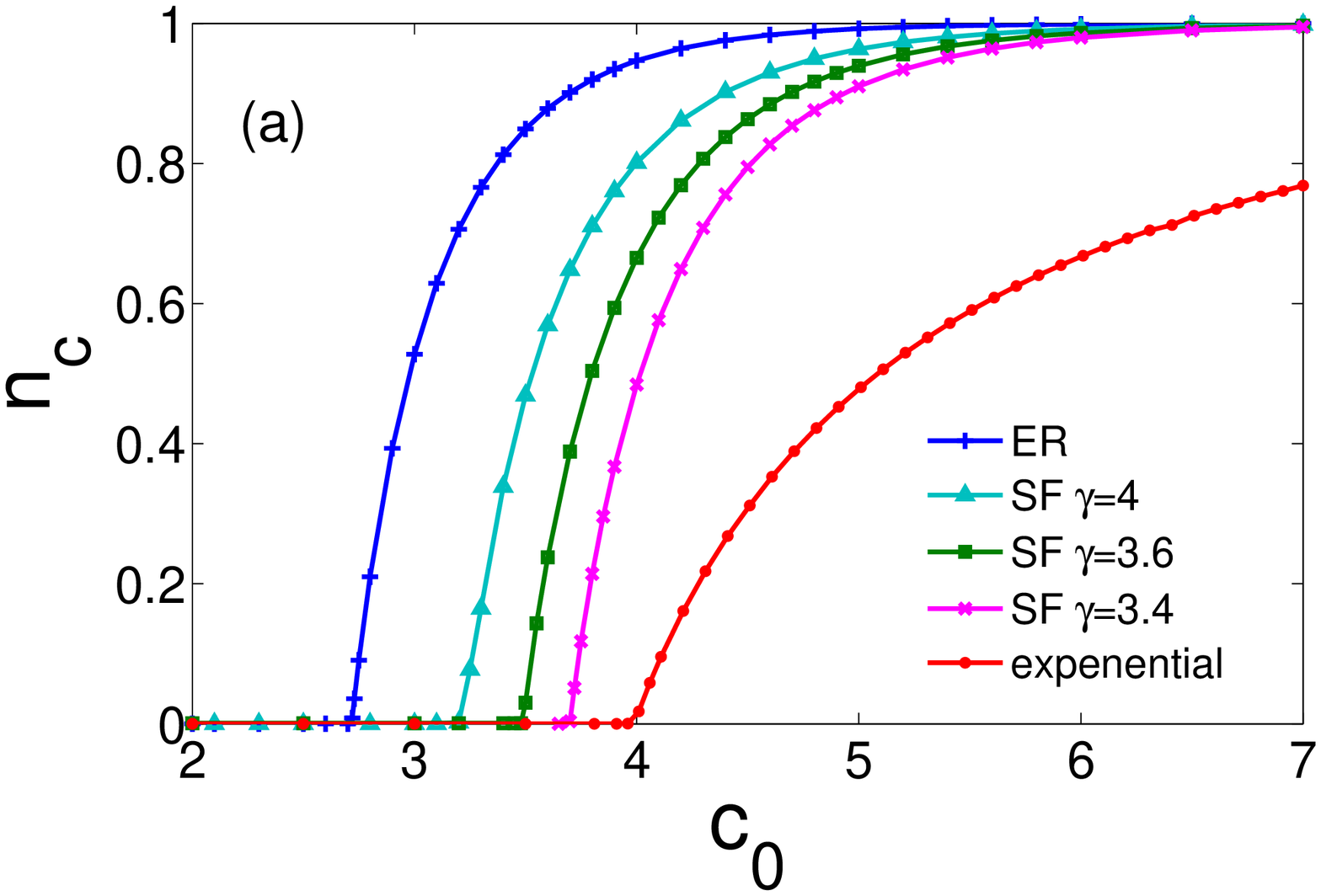}}
\\[-2pt]
\scalebox{0.269}{\includegraphics[angle=0]{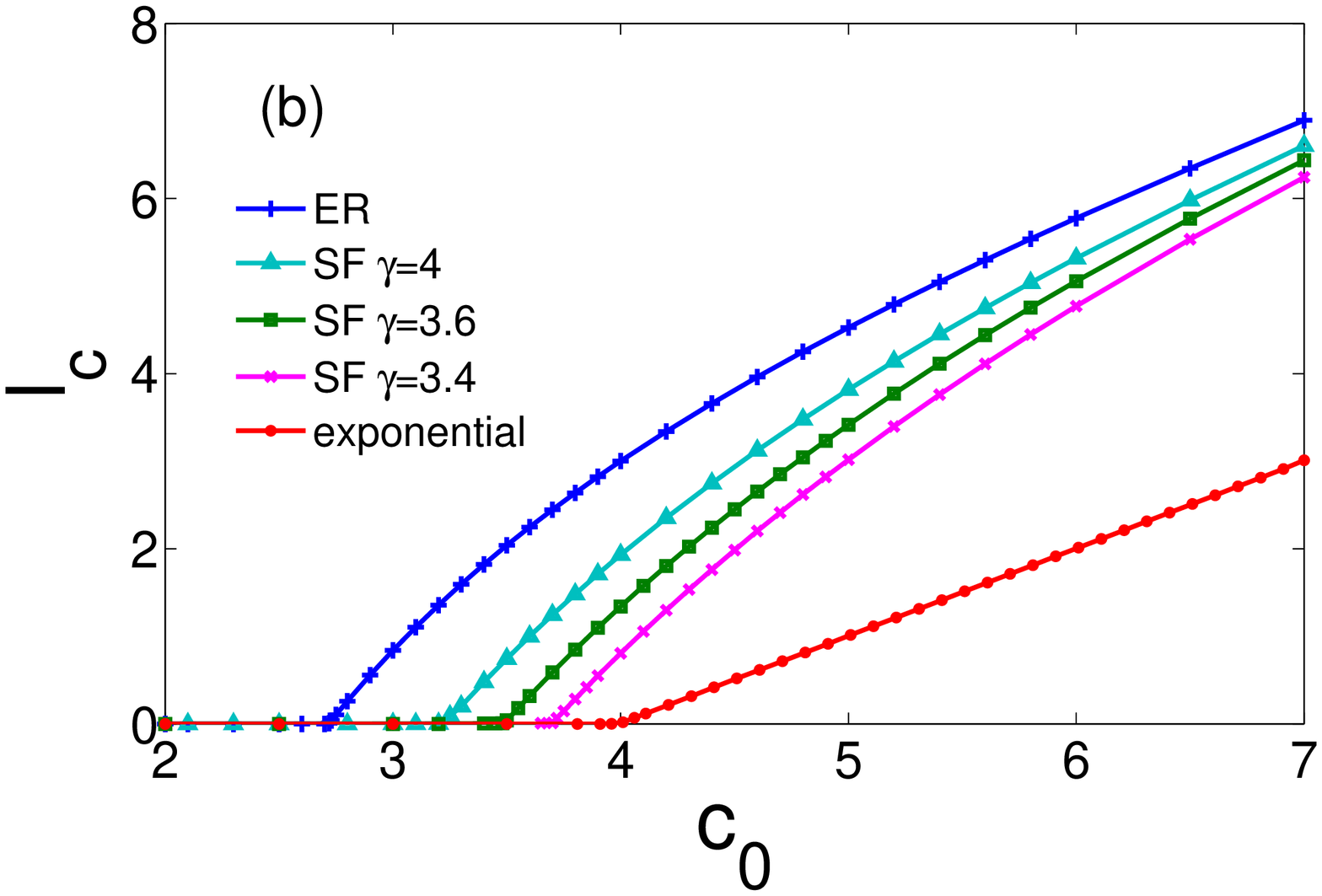}}
\\[-2pt]
\scalebox{0.269}{\includegraphics[angle=0]{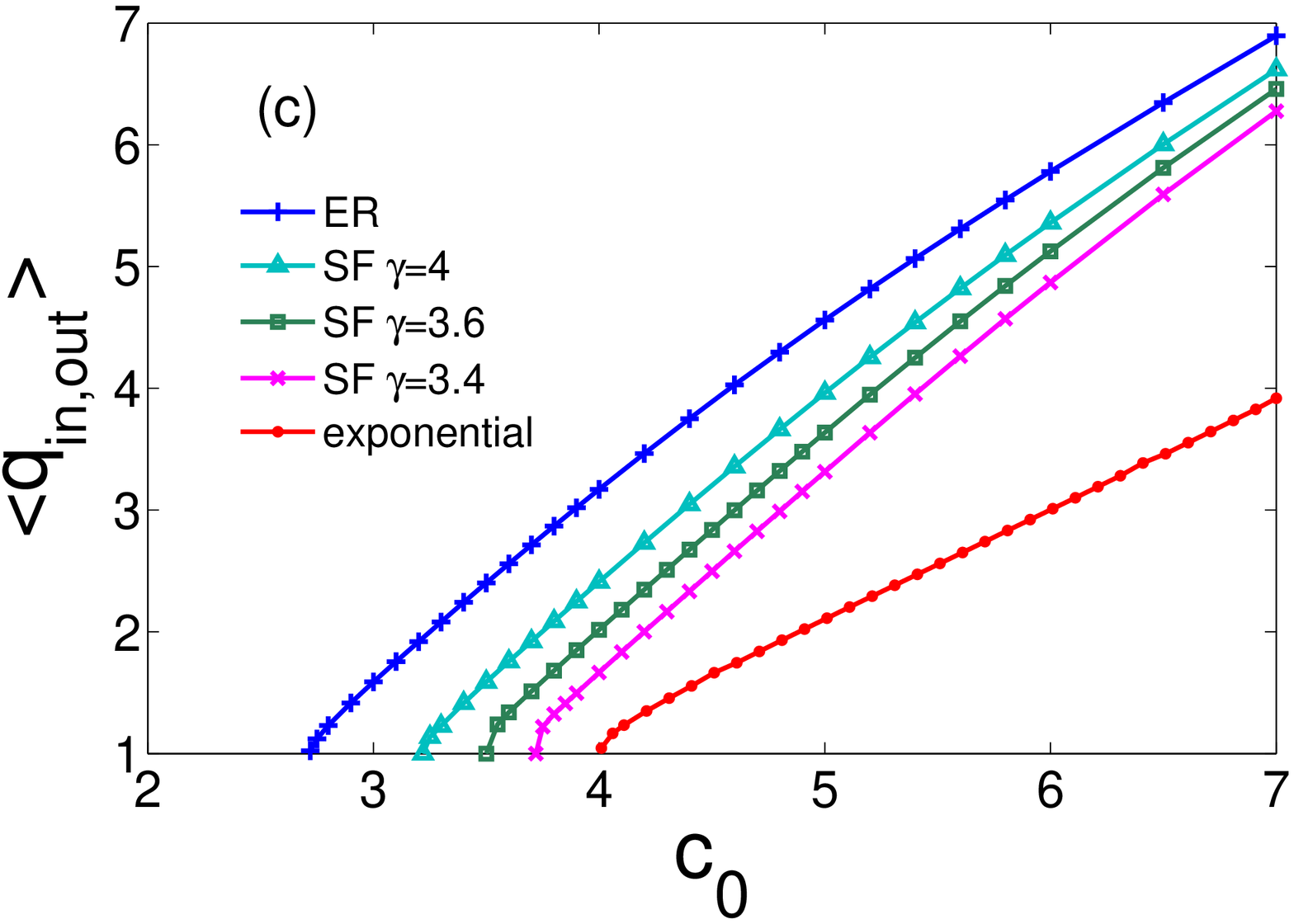}}
\\[-2pt]
\scalebox{0.269}{\includegraphics[angle=0]{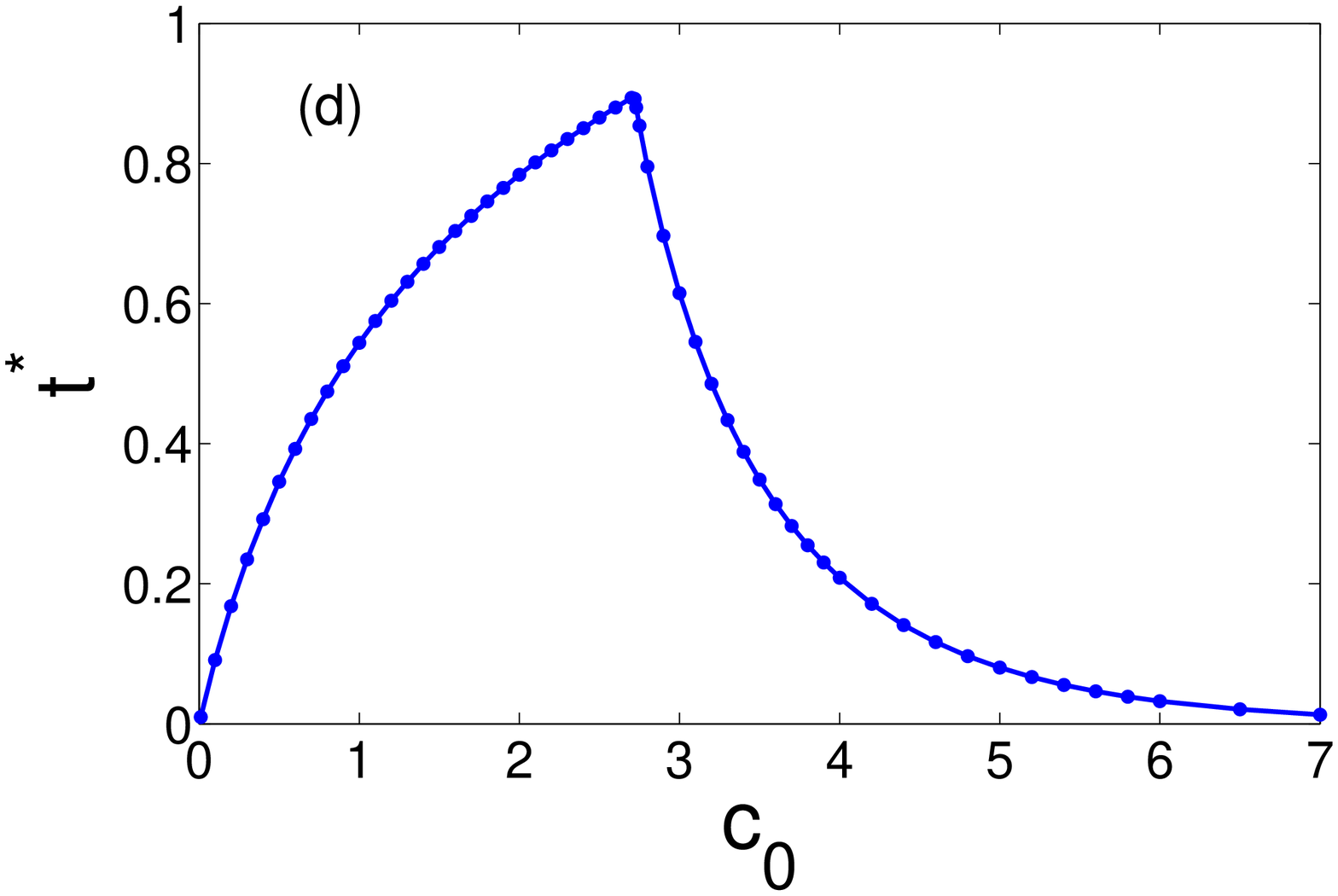}}
\end{center}
\caption{
(Color online) 
(a) The relative size and (b) the normalized number of edges of a core 
vs $c_0$ (the mean in- or out-degree of a vertex in a network before pruning) for different directed networks: 
the directed Erd\H{o}s--R\'enyi graph (ER), the scale-free networks (SF) with the degree distribution exponents $\gamma=4, 3.6, 3.4$, and the network with an exponentially decaying degree distribution.   
(c) The mean in-,out-degree $\langle q_{in} \rangle = \langle q_{out} \rangle = \langle q \rangle/2$ of a vertex in a core 
vs $c_0$ for different directed networks. 
(d) The normalized time $t^*(c_0)$, at which the pruning process finishes,  
vs $c_0$ for the directed Erd\H{o}s--R\'enyi graphs.
}
\label{f6}
\end{figure}



\subsection{$m$-core structures}

As was explained, we can run the algorithm for different values of $m$ and 
so arrive at different $m$-cores. 
The smallest number $m$ is $0$. 
In this case, only leaves with $q_{in}=1,
q_{out}=0$ and $q_{in}=0, q_{out}=1$ are randomly chosen. The leaf
removal algorithm is applied to the network until none of these
leaves remain. The remaining subgraph is $0$-core. 
If
we take $m=1$, we recursively remove the leaves with $q_{in}=1,
q_{out}=0,1$ and $q_{in}=0,1, q_{out}=1$. In this case, we obtain a
smaller subgraph within the $0$-core, namely the $1$-core. 
Similarly, recursively removing 
leaves with degrees 
$q_{in}=1, q_{out}\leq m$ and $q_{in}\leq m, q_{out}=1$ we arrive at 
the $m$-core (should not be confused with ordinary $k$-cores). 
In this way we decompose 
a directed network 
into an onion-like structure of $m$-cores. 
This $m$-core decomposition of the directed networks differs principally from the well-studied $k$-core decomposition. 
($k$-cores are usually defined for undirected graphs, but, in principle, this notion can be generalized to directed networks.) 
In particular, while vertices belonging to inner $k$-cores have a higher mean degree, vertices of inner $m$-cores have a lower mean degree.

To find the structure of the $m$-core of a directed network, we study the evolution 
the degree distribution of leaves during the
leaf removal process. 
In particular, 
we must find when all following probabilities become zero:
$P(0,1, t_m^{*})=P(1,1,t_m^{*})=P(2,1, t_m^{*})=...=P(m,1,t_m^{*})=
P(1,0, t_m^{*})=P(1,1,t_m^{*})=P(1,2, t_m^{*})=...=P(1,m,t_m^{*})=0$. 
Our numerical results 
data for different networks show that $P(1,0)=P(0,1)$ is the last
probability 
to become zero, i.e., the leaves with $q_{in}=1,q_{out}=0$ and $q_{in}=0,q_{out}=1$ are the last to disappear. Hence, to find the time $t_m^*$ at which the pruning process will be completed, we should focus on the evolution of $P(1,0,t)=P(0,1,t)$. 
After we find $t_m^*$, the size
and the number of edges of the $m$-core can be obtained from the following relations: 
\begin{eqnarray}
N_{mc}&=& N[1-P(0,0,t_m^{*})]
,
\label{e9}
\\[5pt]
L_{mc}&=&L(t_m^{*})
.
\label{masterdiredge10}
\end{eqnarray}


\begin{figure}[t]
\begin{center}
\scalebox{0.30}{\includegraphics[angle=0]{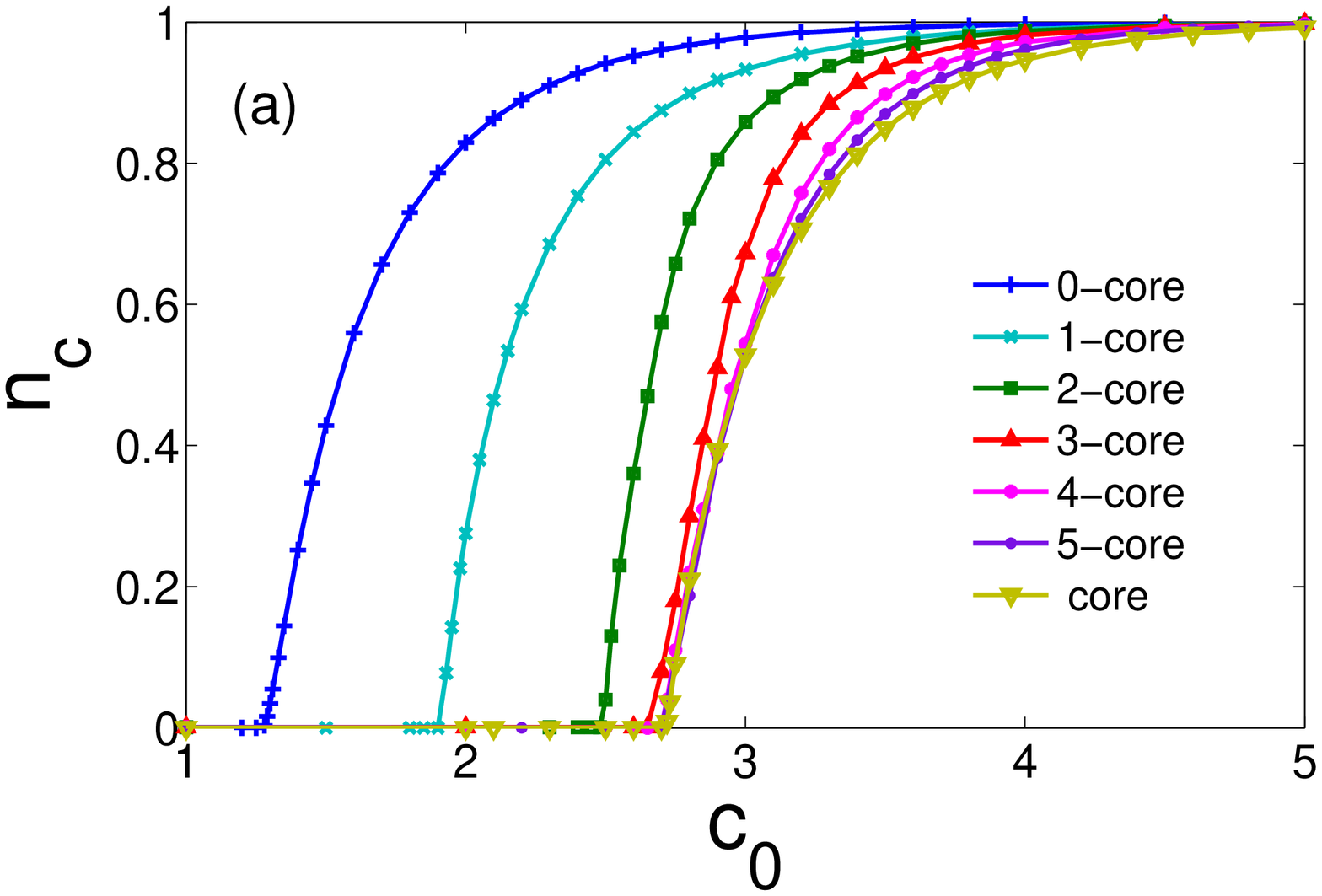}}
\scalebox{0.30}{\includegraphics[angle=0]{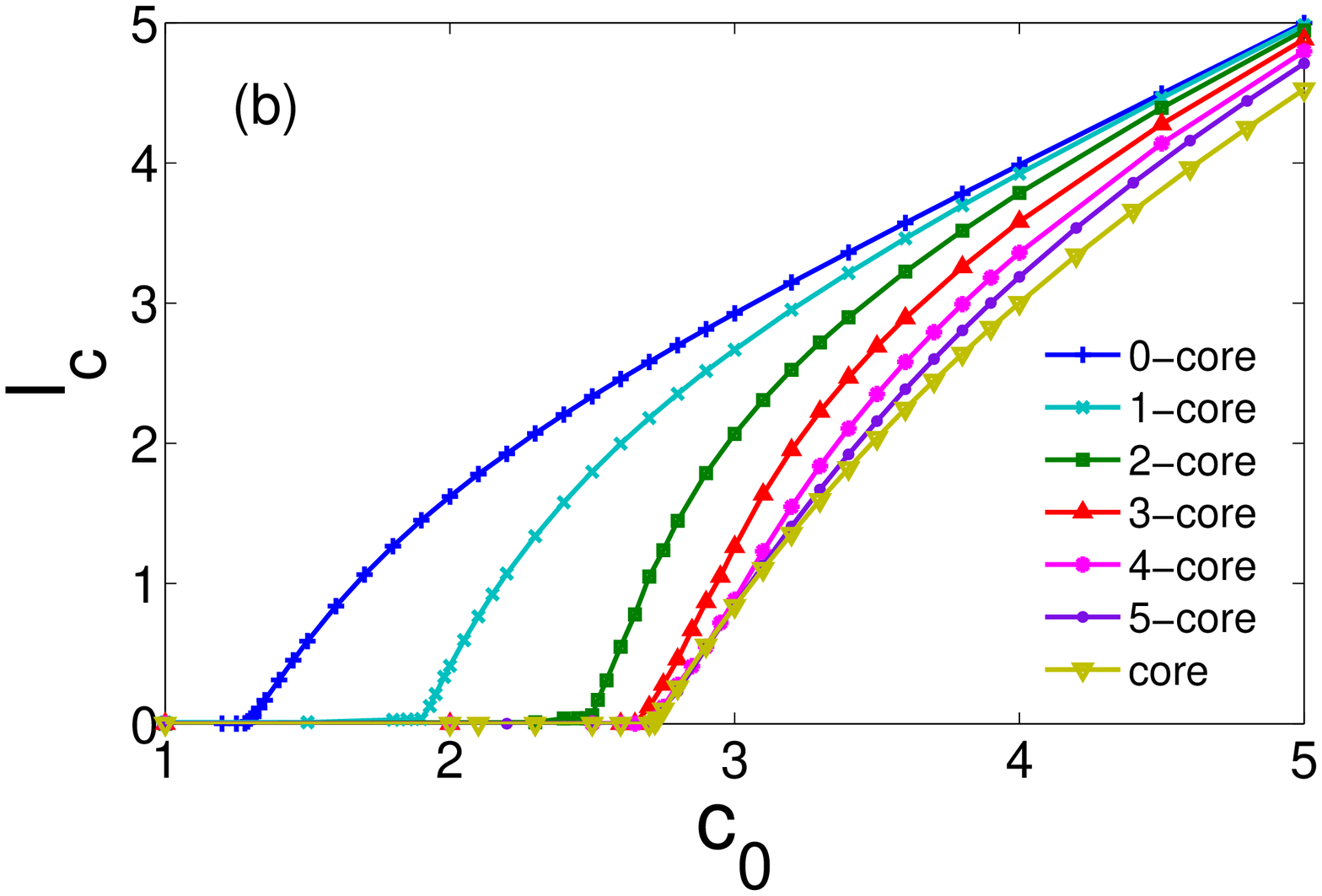}}
\end{center}
\caption{
(Color online) 
$m$-core decomposition for the directed Erd\H{o}s--R\'enyi graph. 
(a) The relative sizes and (b) the normalized numbers of edges of  $m$-cores 
vs $c_0$ (which is the mean in- or out-degree of a vertex in a network). 
}
\label{f7}
\end{figure}


At first, let us study removing of all possible leaves from a directed network, which produces the main core of the network (or simply a core). 
Note that because in our leaf removal algorithm, in- and out-leaves are chosen with equal probability, we can consider only symmetric in,out-degree distributions satisfying $P(q_{in},q_{out}) = P(q_{out},q_{in})$. 
We study networks in which in- and out-degrees of vertices are uncorrelated, i.e., $P(q_{in},q_{out}) = P(q_{in})P(q_{out})$. 
We
apply the algorithm to a few directed uncorrelated networks 
with various degree distributions, namely, (i) a directed Erd\H{o}s--R\'enyi graph in which both $P(q_{in})$ and $P(q_{out})$ are Poissonian, i.e., $P(q) = e^{-c_{0}}c_{0}^{q}/q!$, (ii) 
scale-free networks with different values of the degree distribution exponent, in which both $P(q_{in})$ and $P(q_{out})$ are distributed as $P(q) =[\frac{c_0(\gamma-2)}{2(\gamma-1)}]^{\gamma-1}\Gamma(q-\gamma+1,\frac{c_0(\gamma-2)}{2(\gamma-1)})/\Gamma(q+1) \cong q^{-\gamma}$,   
and (iii) a directed network for which both $P(q_{in})$ and $P(q_{out})$ follow the same exponential law. 
In the leaf removal algorithm, which we perform, in- and out-leaves are chosen with equal probability, and so the evolving in,out-degree distribution remain symmetric during the entire removal process.  
The results, namely, the relative sizes $n_{c}$, and the normalized numbers of edges, $l_{c}$, of the cores in these networks are shown in Fig.~\ref{f6}. 
For the directed networks, $c_{0}$ denotes the mean in-degree, which is equal to the mean
out-degree (half of the total mean degree). 
Comparing Figs.~\ref{f2}(a,b) and \ref{f6}(a,b) one can see that while the critical values of $c_0$ in undirected and directed networks coincide, 
the dependencies $n_c(c_0)$ and $l_c(c_0)$ in undirected and directed networks are essentially different from each other.   
This difference indicates that the problem of cores of directed networks cannot be reduced to that for undirected nets. 

Figure~\ref{f6}(c) demonstrates the dependence of the mean in- or out-degree $\langle q_{in} \rangle = \langle q_{out} \rangle = \langle q \rangle/2$ of a vertex in a core on $c_0$, which is here the mean in- or out-degree of a vertex in a network. We observe that at the critical point, $\langle q_{in,out} \rangle =1$, which indicates the long-loop structure of the core at its birth point. 

Figure~\ref{f6}(d) shows the dependence of the normalized time $t^*$, at which the leaf removal process finishes,   
on $c_0$ for the directed Erd\H{o}s--R\'enyi graphs.
Interestingly, after rescaling, the dependence $t^*(c_0)$ for these directed graphs turns out to be close to the corresponding dependence [see Fig.~\ref{f2}(d)] for their undirected counterparts. Similarly to undirected networks, below the critical point, $t^*$ coincides with the normalized size of the maximum matching.  
Above the critical point, 
$t^*$ provides the normalized size of 
the part of the maximum matching situated outside of the core. 

In Ref.~\cite{control}, the core was found from numerical simulations of the directed version of the Gilbert ($G_{n,p}$) random graphs of $10^4$ vertices, which are equivalent to the corresponding Erd\H{o}s--R\'enyi random graph in the infinite network size limit, and for directed scale-free networks (static model) with the same degree distributions as ours. 
We compared the dependence $n_c(c_0)$ which we obtained for the directed Erd\H{o}s--R\'enyi graph and for the directed scale-free network with $\gamma=4$, see Fig.~\ref{f6}(a), with those in Figs.~S9(b) and (d) from the Supplementary material of Ref.~\cite{control} and observed a close agreement. 

Finally we study the leave removal process involving only leaves with in-degree or out-degree
less than or equal to $m$, which generates $m$-cores. 
There is some maximum value of $m$ at which the corresponding $m$-core coincides with the main core. This number should depend on $c_0$. 
To demonstrate the $m$-core decomposition we use a directed Erd\H{o}s--R\'enyi graph, for which the number of equations in Eq.~(\ref{masterdir}) can be truncated to a small number due to the rapidly decaying Poisson degree distribution. 
We study the range of $c_0$ between $0$ and $5$, choosing $q_{in,
max}=q_{out,max}=20$, and solve numerically $21^2=441$ rate equations for each $m$ fixing the parameter $c_0$. 
Figure~\ref{f7} shows 
the relative size and the normalized 
number of edges for each of $m$-cores. 
The most right curves in this figure show 
$n_{c}(c_0)$ and $l_{c}(c_0)$ for the main core. 
One can see that in the range $e<c_0<3.0$, the $5$-core coincides with the main core. 
Note that each of $m$-cores emerges continuously. These phase transitions are of the second order, and, in the critical region, the $m$-core's size is proportional to the deviation of $c_0$ from a critical point.   

\section{Conclusions}

In the present paper we applied the rate equation approach to the problem of cores in directed networks. The directedness of edges in these networks results in a wide diversity of leaves, which makes this problem essentially more complex than for undirected networks. We developed the leave removal algorithm involving specific leaves and defined the resulting set of $m$-cores forming an onion-like structure 
with smaller and smaller $m$-cores for higher $m$. Studying the evolution of the degree distribution of various networks during 
this algorithm, we obtained the basic characteristics of all these cores and the critical points. We suggest that the equations, which we derived for uncorrelated networks, are asymptotically exact. 
We checked that the result of the application of this approach to the undirected Erd\H{o}s--R\'enyi graph completely agrees with the exact result obtained by the generating function technique. We also found that our results for directed networks 
are in a close agreement with those obtained in simulations.  

We explained that the $m$-leaf and $m$-core notions arise naturally in directed graphs. The open questions however, what is behind these notions, how can one use them, which insight into the structure of a directed network and processes taking place on it can be obtained from the $m$-core decomposition?  
Note that at any given $m$, the $m$-leaf removal process generates matchings. With increasing $m$, the generated matchings become closer and closer to a maximum matching, which is in turn related to the maximum flow through a network \cite{Kleinberg:kt-book05}. When a core emerges, the number of matchings increases crucially.
We suggest that this may 
increase the robustness of flows through a network against removal of random vertices or edges 
within the core, assuming that a large number of matchings provides many options for a flow to bypass the introduced obstacles. Then 
larger flows (large $m$ in the $m$-leaf removal process) should be robust against removal of vertices or edges from $m$-cores that are deeper in the onion structure.  

For the numerical solution of the rate equations we used standard programs for nonlinear differential equations, and, 
for a given finite $q_{\text{max}}$, our results, in principle, could be obtained with any desired precision. We did not however study scale-free networks with degree distributions exponent $\gamma$ smaller than, approximately, $3$ because of the too large number of the rate equations which we had to solve numerically. The case of $\gamma<3$ is a challenge for future work. 

In summary, we have found an effective way to find the complex structure of cores in directed networks. We suggest that, similarly to cores in undirected graphs,  the understanding of cores in directed networks and the proposed $m$-core decomposition 
may shed light on a 
range of phenomena in networks.   

\begin{acknowledgments}
This work was partially supported by the FCT
projects PTDC:
FIS/108476/2008, SAU-NEU/103904/2008, MAT/114515/2009, and PEst-C/CTM/LA0025/2011, and FET IP Project MULTIPLEX 317532.
We thank Y.-Y.~Liu for useful discussions. 
\end{acknowledgments}

\end{document}